\documentclass[12pt]{article}
\usepackage[latin1]{inputenc}
\usepackage{amsfonts}
\usepackage{graphicx}
\usepackage{amsthm}
\usepackage{amsmath}
\usepackage{fancyhdr}
\usepackage{enumerate}
\usepackage{tikz}
\usetikzlibrary{decorations.pathreplacing}
\usepackage{tkz-graph}
\usetikzlibrary{shapes.geometric}
\usetikzlibrary{shapes.multipart}
\usetikzlibrary{arrows,topaths}

\usepackage[agsm,abbr]{harvard}

\title{Inferring structure in bipartite networks using the latent blockmodel and exact ICL}

\author{Jason Wyse$^*$, Nial Friel$^{\dagger}$ and Pierre Latouche$^{\ddag}$
\\
{\small $^{*}$ School of Computer Science and Statistics, Trinity College Dublin, Ireland.}\\
{\small $^{\dagger}$ School of Mathematical Sciences and Insight: The National Centre for Big Data Analytics,}\\
{\small  University College Dublin, Ireland}\\
{\small $^{\ddag}$ Laboratoire SAMM, Universit{\'e} Paris 1 Panth{\'e}on-Sorbonne,} \\{\small 90 rue de Tolbiac, F-75634 Paris Cedex 13, France }}

\date{}

\begin{document}


\maketitle

\newcommand{\ctau}{\hbox{\it$\tau$}}
\newcommand{\cov}{\mbox{cov}}
\newcommand{\diag}{\mbox{diag}}
\newcommand{\Var}{\mbox{var}}
\newcommand{\otherwise}{\mbox{otherwise}}
\newcommand{\boldtheta}{\mbox{\boldmath{$\theta$}}}
\newcommand{\boldlambda}{\mbox{\boldmath{$\lambda$}}}
\newcommand{\boldLambda}{\mbox{\boldmath{$\Lambda$}}}
\newcommand{\boldbeta}{\mbox{\boldmath{$\beta$}}}
\newcommand{\boldmu}{\mbox{\boldmath{$\mu$}}}
\newcommand{\boldvarepsilon}{\mbox{\boldmath{$\varepsilon$}}}
\newcommand{\boldOmega}{\mbox{\boldmath{$\Omega$}}}
\newcommand{\boldSigma}{\mbox{\boldmath{$\Sigma$}}}
\newcommand{\boldrho}{\mbox{\boldmath{$\rho$}}}
\newcommand{\boldomega}{\mbox{\boldmath{$\omega$}}}
\newcommand{\boldphi}{\mbox{\boldmath{$\phi$}}}
\newcommand{\mle}{\mbox{m.l.e.}}
\newcommand{\iid}{\mbox{i.i.d.}}
\newcommand{\Like}{\mathcal{L}}
\newcommand{\normal}{\mbox{N}}
\newcommand{\by}{\mathbf{y}}
\newcommand{\be}{\mathbf{e}}
\newcommand{\bc}{\mathbf{c}}
\newcommand{\bz}{\mathbf{z}}
\newcommand{\bw}{\mathbf{w}}
\newcommand{\bu}{\mathbf{u}}
\newcommand{\bX}{\mathbf{X}}
\newcommand{\bx}{\mathbf{x}}
\newcommand{\bV}{\mathbf{V}}
\newcommand{\diff}{\mbox{d}}
\newcommand{\ICL}{\mbox{ICL}}
\newcommand{\transpose}{\mbox{\tiny T}}

\begin{abstract}
We consider the task of simultaneous clustering of the two node sets involved in a bipartite network. The approach we adopt is based on use of the exact integrated complete likelihood for the latent blockmodel. Using this allows one to infer the number of clusters as well as cluster memberships using a greedy search. This gives a model-based clustering of the node sets. Experiments on simulated bipartite network data show that the greedy search approach is vastly more scalable than competing Markov chain Monte Carlo based methods. Application to a number of real observed bipartite networks demonstrate the algorithms discussed.  
\end{abstract}

\section{Introduction}


Bipartite networks are  those containing two types of nodes,
say types  $A$ and $B$. Nodes  of type $A$  may be linked to  nodes of
type $B$, and \emph{vice versa}, but links between two nodes of the same type are not considered. There are many real life networks that can be naturally viewed in this way. 
Take for example relational networks where a user rates a movie. A user is a member of node type $A$ and node type $B$ represents the movies. One may ask a number of 
quantitative questions in such a situation. Can users be grouped by the types of movies they watch and rate? How many substantive genres of movies are defined by users?

Bipartite or two-mode networks have seen much attention in the social networks and machine learning literature, see for example~\citeasnoun{Borgatti97},~\citeasnoun{Doreian04},~\citeasnoun{Brusco11},~\citeasnoun{Brusco13},
~\citeasnoun{Doreian13}.~\citeasnoun{Rohe12} discuss how their stochastic co-blockmodel may be extended to a bipartite setting.   
The reason for this high level of interest is due to their wide applicability and the fact that it is often very natural and fruitful to model interactions between node sets in this way. Clustering or 
partitioning the node sets simultaneously can reveal structure and give considerable insight into the entities in the network. Such insights may be allusive to the more classical 
network measures \cite{Wasserman94}, some of which have been adapted from the classical literature to the bipartite or two mode situation (for example, the clustering coefficients of \citeasnoun{Opsahl13}). For this reason much attention is focused on clustering or grouping the node sets in tandem; this practice is referred to using many terms in the literature: bi-clustering, co-clustering, block-clustering, two-mode blockmodelling and others. One of the pioneering papers of this area was that of~\citeasnoun{Hartigan72}.~\citeasnoun{Flynn12} examine asymptotic theoretical guarantees for bi-clustering. 

Approaches to clustering or blockmodelling of two-mode networks fall into two classes. The first of these is often called \emph{deterministic}, whereby the clustering of the network (or adjacency matrix) is obtained by minimizing an objective function which measures discrepancy from an ideal block structure. Examples of this are the work of~\citeasnoun{Doreian04},~\citeasnoun{Brusco06},~\citeasnoun{Brusco11},~\citeasnoun{Brusco13} and ~\citeasnoun{Doreian13}. The second type of approach is \emph{stochastic}. In stochastic blockmodelling procedures, one assumes that the probability of links between the node sets in the network can be modelled by a parameterized distribution. These parameters are usually estimated (learned) and then used as a representative embodiment of the true network linking behaviour. The stochastic approach may also be referred to as model-based; that is, a statistical model is used for links in the network. Examples include the work of~\citeasnoun{Govaert95},~\citeasnoun{Govaert96},~\citeasnoun{Govaert03},~\citeasnoun{Govaert05},~\citeasnoun{Govaert08},~\citeasnoun{Rohe12},~\citeasnoun{Wyse12} and ~\citeasnoun{Keribin13}. 


This paper is concerned with the stochastic approach to blockmodelling of two-mode  networks. In particular, we take the latent blockmodel (LBM) developed in a series of papers by G{\'e}rard Govaert and Mohamed Nadif; ~\citeasnoun{Govaert95},~\citeasnoun{Govaert96},~\citeasnoun{Govaert03},~\citeasnoun{Govaert05},~\citeasnoun{Govaert08}. We consider this model as applied in the context of bipartite networks. This is a desirable model, as it provides a 
model-based clustering  of both  node sets  and has  richer modelling
capability  than only  absence/presence  data for  ties between  nodes
should this information be observed. The LBM is based around an intuitive generative structure as outlined in Section~\ref{sec:lbm}. We point out that the LBM is a different model than the stochastic blockmodel (SBM), as in, for example,~\citeasnoun{Nowicki01}. The LBM operates on items and objects as opposed to the SBM which focusses on modelling interactions between items. 

In the LBM, the posterior
distribution over the latent label vectors, given the model parameters
and   observed  data,   cannot  be   factorized  due   to  conditional
dependency.  Therefore standard  optimization techniques  such  as the
expectation  maximization (EM) algorithm   cannot  be  used  directly  for
clustering.   To tackle  this issue,  approximation  methods
like variational EM \cite{Govaert08} or stochastic EM \cite{keribin10} have been proposed. Moreover, in practice, the
numbers of  clusters in each node  set have to be  estimated. This has
led  to treatements  such as  the one  by~\citeasnoun{Wyse12}  who use
Markov  chain Monte Carlo  (MCMC) to  do inference  for the  number of
clusters and the members of the  nodes to the groups. We also refer to
the recent work of \citeasnoun{Keribin2012} and \citeasnoun{Keribin13}
who  relied on  model selection  criteria  to estimate  the number  of
clusters. 

 Unlike
 ~\citeasnoun{Govaert08},~\citeasnoun{Keribin2012},~\citeasnoun{Keribin13},
 our approach allows the number of clusters in both node sets to be estimated while simultaneously partitioning the nodes. This is based on a clustering criterion termed the exact integrated complete likelihood (ICL), and a method to search over partitions of the nodes. The main ideas of using the exact ICL come from~\citeasnoun{Come13} who use this in estimation of the SBM of~\citeasnoun{Nowicki01}. Our work 
can be seen as somewhat complementary to the work in~\citeasnoun{Wyse12}, but there are many advantages to using the framework presented here, similar to the SBM in~\citeasnoun{Come13}. Firstly, it is more scalable than the MCMC approach and secondly we do not have to worry about the mixing rates of the MCMC algorithm in larger settings. One drawback of our new approach is that it does not provide a joint posterior distribution for the number of clusters in both node sets.

The     remainder     of     the     paper     is     organised     as
follows.  Section~\ref{sec:sec2} introduces  ideas of  blockmodels for
bipartite   networks  (Section~\ref{sec:bipart})   and   the  LBM   of
~\citeasnoun{Govaert08}  (Section~\ref{sec:lbm})   and  discusses  its
relevance for this context. Section~\ref{sec:lbm_est} reviews existing
estimation techniques  for the LBM and discusses  their advantages and
limitations.  In  Section~\ref{sec:sec3} we  introduce  the exact  ICL
along with a greedy algorithm for inference purposes. We also draw strong parallels with the approach of~\citeasnoun{Wyse12} in this section, and highlight the sensible and pragmatic nature of the algorithm. The computational complexity of the algorithm is discussed in Section~\ref{sec:complex}. In Section~\ref{sec:sec4} we show how large computational savings can be made over a naive implementation of this algorithm. Section~\ref{sec:simstud} compares the proposed approach with the competing MCMC algorithm through a simulation study. Section~\ref{sec:sec5} applies the approach to a number of datasets. We conclude with a discussion. 

\section{Bipartite networks and the latent blockmodel} \label{sec:sec2}

\subsection{Bipartite networks} \label{sec:bipart}

Bipartite  networks consist of  possible ties  between members  of two
different node sets.  Let the node sets be $A$ and  $B$. There are $N$
nodes  in   $A$,  $A   =  \{a_1,\dots,a_N\}$  and   $M$  in   $B$,  $B
=\{b_1,\dots,b_M\}$.   We   use   the   terms  node   type   and   set
interchangeably. Node  $i$ from $A$ may  have a tie with  node $j$ from
$B$, this tie,  if it exists, is undirected.  Alternatively, a tie may
not exist. Either way, this information on the {\it linking attribute}
between the two nodes is contained in the observation $y_{ij}$. We record all the observed information about the network in an $N \times M$ adjacency matrix
\[
Y = \left(\begin{array}{ccc}
y_{11} & \dots &y_{1M}\\
\vdots & \ddots & \vdots \\
y_{N1} & \dots & y_{NM}
\end{array} \right)
\] 
having a row for each node from $A$ and a column for each in $B$. The problem of primary focus for us is the following. Can we group or partition the $A$ nodes and the $B$ nodes simultaneously to reveal subgroups or subsets of the $A$ nodes that have linking attributes of similar nature to subgroups of the $B$ nodes? We use the term ``similar nature'' here to highlight that we approach this problem by modelling properties of the linking attribute which could be binary in nature, categorical ($>2$ categories), a count or an observation with a continuous value. 

Suppose a grouping of the node sets 
into subsets like that mentioned above does exist. The nodes in $A$ are partitioned into $K$ and those in $B$ into $G$ disjoint subsets
\[
A = \bigcup_{k=1}^{K} A_k,\quad A_k \cap A_l = \emptyset, \forall k \ne l \quad\mbox{ and }\quad
B = \bigcup_{g=1}^G B_g ,\quad B_g \cap B_h = \emptyset, \forall g \ne h.
\]
Then the nodes in set $A_k$ will be seen as having similar natured linking attributes to their respective nodes in set $B_g$. We suppose now that we model linking attributes using some parametric distribution $p(y_{ij}|\boldtheta)$ for the $y_{ij}$. In order to capture differences in the nature of linking attributes between the different subsets, we allow this distribution to have a different valued parameter $\boldtheta_{kg}$ for each subset pairing $A_k$ and $B_g$, so that if node $i$ is in $A_k$ and node $j$ is in $B_g$, the density $y_{ij}$ is $p(y_{ij}|\boldtheta_{kg})$.

\subsection{Latent blockmodel} \label{sec:lbm}

The LBM was introduced in~\citeasnoun{Govaert08} as a means to provide
a  concise summary  of a  large  data matrix.  In order  to index  the
different  possible partitionings  of $A$  and  $B$ into  $K$ and  $G$
disjoint  subsets  (clusters) we  introduce  respective label  vectors
$\bc$ and $\bw$. These are such that $c_i = k$ if node $i \in A_k$ and
similarly  $w_j =  g$ if  $j  \in B_g$.  Using the  model for  linking
attributes,  the probability  of observing  the adjacency  $Y$  can be
written down, if we assume we know the partitioning of the nodes, $\mbox{i.e.}$ that know $\bc$ and $\bw$
\[
p(Y|\bc,\bw,\Theta,K,G) = \prod_{k=1}^K \prod_{g=1}^G \prod_{i:c_i = k} \prod_{j:w_j=g} p(y_{ij}|\boldtheta_{kg}).
\]
Here $\Theta$ denotes the collection of $\boldtheta_{kg}$. 

Of course,  in practice $\bc$ and  $\bw$ will not be  known. Attach to
each     clustering    of     nodes     $\bc,\bw$    a     probability
$p(\bc,\bw|\Phi,K,G)$, depending on  some hyperparameters $\Phi$. This
allows the density of the adjacency matrix $Y$ to be written down as a mixture model. Let $\mathcal{C}_K$ and $\mathcal{W}_G$ be collections containing all possible clusterings of $A$ into $K$ groups and $B$ into $G$ groups respectively. Then
\begin{equation}
p(Y|\Theta,\Phi,K,G) = \sum_{(\bc,\bw)\in \mathcal{C}_K \times \mathcal{W}_G} p(\bc,\bw|\Phi,K,G) p(Y|\bc,\bw,\Theta,K,G). \label{eq:py}
\end{equation}
Here the mixture is over all clusterings of the nodes into $K$ and $G$ groups. The $p(\bc,\bw|\Phi,K,G)$ terms represent probabilities of particular clusterings being generated parameterized by $\Phi$. 

Suppose that, {\it a priori}, no information exists on the joint clustering of nodes in sets $A$ and $B$. Then it is reasonable to assume that 
\[
p(\bc,\bw|\Phi,K,G) = p(\bc|\Phi,K)p(\bw|\Phi,G).
\]
Having made this assumption, LBM assumes further that there are weights associated with each subset $A_k$ and $B_g$, such that 
$
\Pr\{c_i = k|K\} = \omega_k \mbox{ and } \Pr\{w_j = g|G\} = \rho_g
$. This defines a multinomial distribution for the node labels, where the weights sum to unity: $\sum_{k=1}^K \omega_k = 1$, $\sum_{g=1}^G \rho_g = 1$. The parameters $\Phi$ represent the weight vectors $\boldomega,\boldrho$. Thus we can write 
\[
p(\bc,\bw|\Phi,K,G) = \left(\prod_{k=1}^K \omega_k^{N_k} \right) \left( \prod_{g=1}^G \rho_g^{M_g} \right)
\] 
where $N_k = |A_k|$ and $M_g = |B_g|$.

Effectively the LBM defines a probability distribution over clustering of the node sets $A$ and $B$ into $K$ and $G$ disjoint subsets.  Examining equation~(\ref{eq:py}) one sees that the sum is over $K^N G^M$ terms, which is clearly intractable for even moderate $N$ and $M$. In practice, we are interested in finding optimal or near optimal clusterings, corresponding to large values of $p(\bc,\bw|\Phi,K,G)p(Y|\bc,\bw,\Theta,K,G)$ in (\ref{eq:py}) which contribute the most to the sum. It is thus usual to rephrase the problem into one which optimises $p(\bc,\bw|\Phi,K,G)p(Y|\bc,\bw,\Theta,K,G)$ jointly over the clusterings $(\bc,\bw)$ and parameters $\Phi,\Theta$ in some way. Something to note here is that the particular cluster configurations are conditional on the value of $K$ and $G$ being known, something which will rarely be the case in real applications.

\subsection{Estimation of the LBM} \label{sec:lbm_est}

Estimation  of  the  LBM   can  be  performed  either  through  hybrid
variational      and      stochastic      EM      type      algorithms
\cite{Govaert08,keribin10}    or     through    Bayesian    estimation
(\citeasnoun{vanDijk09},~\citeasnoun{Wyse12}). We  refer the reader to
Section 2.1.1 of~\citeasnoun{Wyse12} for  a review of the former. Here
we  focus on  Bayesian  estimation as  it  is most  relevant for  what
follows. The one remark to be made about EM type estimation of the LBM
however is that it conditions on the values of $K$ and $G$. As of yet,
there  is no widely  accepted information  criterion for  choosing the
best   values  of   $K$  and   $G$  as   outlined  in   Section  2.1.2
of~\citeasnoun{Wyse12}. \citeasnoun{Keribin2012},~\citeasnoun{Keribin13}
relied  on  a  ICL  criterion  for  model  selection  purposes,  while
\citeasnoun{vanDijk09}   proposed  a   Gibbs   sampler  for   Bayesian
estimation   of  the   latent  block   model  along   with   an  AIC-3
\cite{Bozdogan94} criterion.

~\citeasnoun{Wyse12} circumvented the problem of choosing $K$ and $G$ by including these as unknowns in their Bayesian formulation of LBM. They begin by taking priors on the unknowns in the model, namely, $\Theta,\Phi,K$ and $G$ and writing down the full posterior given the data
\begin{eqnarray}
\pi(K,G,\bc,\bw,\Phi,\Theta|Y)& \propto  & p(\bc,\bw|\Phi,K,G)p(Y|\bc,\bw,\Theta,K,G) \nonumber\\
& & \times \pi(\Theta|K,G)\pi(\Phi|K,G)\pi(K)\pi(G) \label{eq:priors}\\
& = & p(\bc,\bw|\Phi,K,G)\pi(\Phi|K,G) \nonumber\\ & & \times  p(Y|\bc,\bw,\Theta,K,G)\pi(\Theta|K,G) \pi(K)\pi(G)\nonumber
\end{eqnarray}
where the last  three distributions on the right  hand side are priors
for $\Theta,K,G$. Note  here the assumption that, {\it  a priori}, the
number of node clusters $K$ and $G$ of types $A$ and $B$ are independent. If other prior information exists beforehand, this prior assumption can be replaced by one which represents this information. Integrating both sides of the above proportionality relation with respect to $\Phi$ and $\Theta$ returns the joint marginal distribution of $K,G,\bc,\bw$ given the observed network $Y$:
\begin{eqnarray*}
\pi(K,G,\bc,\bw|Y) &\propto & \int p(\bc,\bw|\Phi,K,G) \pi(\Phi|K,G) \, \mathrm{d}\Phi\\
 & & \times \int p(Y|\bc,\bw,\Theta,K,G)\pi(\Theta|K,G) \,\mathrm{d}\Theta\\
 & &\times \pi(K) \,\pi(G)\\
 & = & \pi(\bc,\bw|K,G) \pi(Y|\bc,\bw,K,G) \pi(K) \pi(G).
\end{eqnarray*}
The key observation in~\citeasnoun{Wyse12} is that the quantities
\begin{eqnarray*}
\pi(\bc,\bw|K,G)& =& \int p(\bc,\bw|\Phi,K,G) \pi(\Phi|K,G) \, \mathrm{d}\Phi\\
\pi(Y|\bc,\bw,K,G) & = & \int p(Y|\bc,\bw,\Theta,K,G)\pi(\Theta|K,G) \,\mathrm{d}\Theta
\end{eqnarray*}
can be obtained exactly (analytically) using relatively standard prior assumptions. For example $\pi(\bc,\bw|K,G)$ can be computed exactly by assuming independent Dirichlet priors on the weights vectors
\[
\pi(\Phi|K,G) = \mathrm{Dir}_{K}(\boldomega; \alpha_0,\dots,\alpha_0) \times \mathrm{Dir}_{G}(\boldrho; \beta_0,\dots,\beta_0)
\]
where $\mathrm{Dir}_p$ represents the density of the Dirichlet distribution on the $p-1$ dimensional simplex. In this case
\[
\pi(\bc,\bw|K,G) = \frac{\Gamma\{\alpha_0 K\}}{\Gamma\{\alpha_0\}^K} \frac{\prod_{k=1}^K \Gamma\{N_k + \alpha_0\}}{\Gamma\{N + \alpha_0 K\}} \times \frac{\Gamma\{\beta_0 G\} }{\Gamma\{\beta_0\}^G}\frac{\prod_{g=1}^G\Gamma\{M_g + \beta_0\}}{\Gamma\{M + \beta_0 G\}}
\] 
where $\Gamma(\cdot)$ is the gamma function. See~\citeasnoun{Wyse12} for further details. Furthermore $\pi(Y|\bc,\bw,K,G)$ can be computed exactly if we make the following assumptions. Assume that the prior for $\Theta$ can be expressed as independent priors for the $\boldtheta_{kg}$,
\[
\pi(\Theta|K,G) = \prod_{k=1}^K \prod_{g=1}^G \pi(\boldtheta_{kg}).
\]
We can compute $\pi(Y|\bc,\bw,K,G)$ exactly if we can compute
\[
\Lambda_{kg} = \int \pi(\boldtheta_{kg}) \prod_{i:c_i = k} \prod_{j:w_j=g} p(y_{ij}|\boldtheta_{kg})\,\mathrm{d} \boldtheta_{kg}
\]
exactly. This happens when the prior $\pi(\boldtheta_{kg})$ is fully conjugate to $p(y|\theta_{kg})$. There are many widely used and standard situations where this is the case as we discuss in Section~\ref{sec:models}. In our notation we suppress dependence on the values of the hyperparameters, say $\alpha_0, \beta_0$ and the hyperparameters of $\pi(\boldtheta_{kg})$ and similar for brevity. This is to be understood.

With these quantities available exactly, the joint marginal posterior of clusterings and number of clusters of the nodes is given by
\begin{equation}
\pi(K,G,\bc,\bw|Y) \propto \pi(\bc,\bw|K,G)\left( \prod_{k=1}^K \prod_{g=1}^G \Lambda_{kg}\right) \pi(K) \pi(G). \label{eq:post}
\end{equation}
Notice that this posterior is defined over a large discrete model space. If we allow a maximum of $K_{\mathrm{max}}$ and $G_{\max}$ subsets for nodes in $A$ and $B$, then the size of the support of the posterior is $\sum_{k=1}^{K_{\mathrm{max}}}\sum_{g=1}^{G_{\max}} k^N g^M$. 

~\citeasnoun{Wyse12} developed an  MCMC algorithm that generates samples
from (\ref{eq:post}).  The algorithm moves are involved  and we refer
to~\citeasnoun{Wyse12}  for  exact details.  Node  labels are  sampled
using a Gibbs  step and there are more  sophisticated moves for adding
and removing clusters  (that is, changing $K$ and  $G$). The main idea
is   that  iteratively   the  chain   will  sample   high  probability
configurations $(K,G,\bc,\bw)$, so that in the sampler output one gets
a  joint posterior  distribution for  $K$ and  $G$  with corresponding
label vectors. There are upper bounds $K_{\max}$ and $G_{\max}$ placed
on the  number of  groups the nodes  may be partitioned  into. Usually
this will  be conservatively  large. A drawback  of this  algorithm in
general  is  that mixing  of  the  Markov  chain may  disimprove  with
increasing  $M$ and/or $N$,  resulting in  infrequent jumps  to models
with  different  $K$ and  $G$  from the  current  state.  So for  high
dimensional  problems,   one  can   observe  most  of   the  empirical
(approximate)   posterior  mass   centred  on   one  or   two  $(K,G)$
combinations.  However,  the  work  of~\citeasnoun{Wyse12} is  a  step
forward in  estimation of LBMs  as it is  an automatic way  to perform
inference for $K$ and $G$ also, while clustering the nodes, for which no other such approach existed before. 

\section{Exact ICL and greedy ICL algorithm for bipartite networks} \label{sec:sec3}

\subsection{ICL and exact ICL} \label{sec:ex_ICL}

The  integrated completed  likelihood (ICL)  was first  introduced in
~\citeasnoun{Biernacki01}, as a model selection criterion, in the context of Gaussian mixture models. The rationale for using ICL is that in the finite mixture model context (similar to our situation here), analysis is often carried out using a latent label vector which it is difficult to integrate from the model. This latent vector is often termed the allocation vector~\cite{Nobile07} and it provides a clustering of the data points to component densities. For this reason~\citeasnoun{Biernacki01} argue that the evidence for a particular clustering should be taken into account when determining the number of mixture components and thus they focus on the integrated completed data likelihood. So instead of marginalising these labels from the model, ICL includes them as part of the information criterion. The number of components in the mixture (in our case values of $K,G$) which gives the largest ICL is the most supported by the data. For full details see~\citeasnoun{Biernacki01}. \citeasnoun{Keribin13} have used 
ICL in the context of the latent blockmodel to choose the number of clusters, however, it is not the same approach as we adopt here as they do not use the available form of the exact ICL. Instead, their analysis employs a penalty term in the ICL criterion, which we avoid by computing ICL analytically. Using the notation introduced earlier, the ICL can be written as
\[
\mathcal{ICL} = \log\left\{ \int \int p(\bc,\bw|\Phi,K,G)p(Y|\bc,\bw,K,G) \pi(\Theta|K,G)\pi(\Phi|K,G) \, \mathrm{d} \Theta \, \mathrm{d} \Phi\right\}.
\]
Typically this quantity cannot be computed exactly and so ICL is usually approximated using a high probability configuration and a penalty term. In Section~\ref{sec:lbm_est} we encountered a similar expression to $\mathcal{ICL}$ in the joint collapsed posterior of~\citeasnoun{Wyse12}, and we see that we can write
\begin{eqnarray*}
\mathcal{ICL} & = & \log\left\{ \int \int p(\bc,\bw|\Phi,K,G)p(Y|\bc,\bw,K,G) \pi(\Theta|K,G)\pi(\Phi|K,G) \, \mathrm{d} \Theta \, \mathrm{d} \Phi\right\}\\
& = & \log \left\{\left( \int  p(\bc,\bw|\Phi,K,G)\pi(\Phi|K,G)\mathrm{d} \Phi\right) \times \left(\int p(Y|\bc,\bw,K,G) \pi(\Theta|K,G)\mathrm{d} \Theta\right)\right\}\\
& = & \log \left\{\int  p(\bc,\bw|\Phi,K,G)\pi(\Phi|K,G)\mathrm{d} \Phi \right\} + \log \left\{\int p(Y|\bc,\bw,K,G) \pi(\Theta|K,G)\mathrm{d} \Theta \right\}\\
& = & \log \pi(\bc,\bw|K,G) + \log \pi(Y|\bc,\bw,K,G).
\end{eqnarray*}
The conditions for the ICL to be available exactly are the same as those given in Section~\ref{sec:lbm_est}. If this is the case we term this quantity {\it exact} ICL. Now suppose we can compute the exact ICL so that
\[
\mathcal{ICL} = \log \pi(\bc,\bw|K,G) + \sum_{k=1}^K \sum_{g=1}^G \log \Lambda_{kg}.
\] 
Note the similarity between this expression and the log of the right hand side of (\ref{eq:post}). In fact, the only difference is that in (\ref{eq:post}) we also have two log prior terms for $K$ and $G$. Thus, finding the highest ICL is almost identical to finding regions of the support of (\ref{eq:post}) with high posterior mass (bar the prior terms). This is elucidated more in Section~\ref{sec:sensible}. 

\subsection{Linking attribute models} \label{sec:models}

We now turn attention to the types of linking attribute models for which exact ICL can be computed. The requirement for the exact ICL to be available is that 
\[
\Lambda_{kg} = \int \pi(\boldtheta_{kg}) \prod_{i:c_i = k} \prod_{j:w_j=g} p(y_{ij}|\boldtheta_{kg}) \, \mathrm{d}\boldtheta_{kg}
\]
can be computed analytically. $\Lambda_{kg}$ can be referred to as the marginal likelihood for block $(k,g)$, as we pick up nodes from cluster $k$ of $A$ and cluster $g$ of set $B$ and average their likelihood over the prior for the parameter of the linking attributed between the two node subsets. We now give some examples of very commonly used models where the marginal likelihood can be computed exactly.

\subsubsection{Absent/present linking attribute}

The most commonly observed networks will be those where edges are represented by binary indicators when there is a link between a node in set $A$ and a node in set $B$ or not. For example, students being members of university clubs or not. For $i\in  A_k$ and $j \in B_g$ the most natural way to model this is using a Bernoulli random variable with probability of a tie equal to $\theta_{kg}$
\[
p(y_{ij}|\theta_{kg}) = \theta_{kg}^{y_{ij}}(1-\theta_{kg})^{1-y_{ij}}.
\]
A conjugate prior for this data distribution is a beta prior 
\[
\theta_{kg} \sim \mathrm{Beta}(\eta,\eta)
\]
with hyperparameter $\kappa$. Then it can be easily shown that 
\[
\Lambda_{kg} = \frac{\Gamma\{2\eta\}}{\Gamma\{\eta\} \Gamma\{\eta\}} \frac{\Gamma\{N_{kg}^1 + \eta\}\Gamma\{N_k M_g - N_{kg}^1 + \eta\}}{\Gamma\{N_k M_g + 2 \eta\}}
\]
where $N_{kg}^1 = \sum_{i:c_i = k} \sum_{j: w_j=g} \mathrm{I}(y_{ij} = 1)$.

\subsubsection{Multinomial links with Dirichlet prior}

This is a generalization of the previous model for more than two categories. Linking attributes which are categorical arising from $C$ categories could naturally be modelled using the multinomial distribution. This could be useful say if the college years of the students in the university clubs above were also recorded $\mbox{e.g.}$ junior freshman, senior freshman, junior sophister, senior sophister. The distribution is 
\[
p(y_{ij}|\boldtheta_{kg}) = \prod_{l=1}^C \left[\theta_{kg}^{\,l}\right]^{\mathrm{I}(y_{ij} = l)}
\]
where $\theta_{kg}^{\,l}$ is the probability $y_{ij}$ takes category $l$. Taking a symmetric Dirichlet prior with parameter $\zeta$ the block integrated likelihood can be computed exactly as
\[
\Lambda_{kg} = \frac{\Gamma(\zeta C)}{\Gamma(\zeta)^C} \frac{\prod_{l=1}^C \Gamma(N_{kg}^l + \zeta)}{\Gamma(N_k M_g +  \zeta C)}
\]
where $N_{kg}^l = \sum \limits_{i:c_i=k} \sum\limits_{j:w_j=g} \mathrm{I}(y_{ij} = l)$ is the number taking on category $l$ in block $(k,g)$. 

\subsubsection{Poisson model for count links with Gamma prior}

Count data may be modelled using a Poisson distribution with rate $\theta_{kg}$. This could be the number of emails exchanged between students in the university clubs. Taking a $\mathrm{Gamma}(\delta,\gamma)$ prior on $\theta_{kg}$ leads to 
\[
\Lambda_{kg} = \frac{\gamma^{\delta}}{\Gamma(\delta)}\frac{ \Gamma(S_{kg} + \delta)}{(N_k M_g + \gamma)^{S_{kg}+\delta}} \left( \prod_{i:c_i=k} \prod_{j:w_j=g} y_{ij}!\right)^{-1}
\]
where $S_{kg} = \sum\limits_{i:c_i=k} \sum\limits_{j:w_j=g} y_{ij}$ is the sum of the counts in a block. 

\subsubsection{Gaussian model with Gaussian-Gamma prior}

For continuous data, if appropriate one can assume a Gaussian distribution with mean $\mu_{kg}$ and precision $\tau_{kg}$ for observations in a block. This may be the number of hours spent by the university students on club activities. For example, one would expect a sports club to require greater time commitment than, say, a car appreciation club. Taking a $\mathrm{N}(\xi,\{\kappa \tau_{kg}\}^{-1})$ prior for $\mu_{kg}$ conditional on $\tau_{kg}$ and a $\mathrm{Gamma}(\gamma/2,\delta/2)$ prior on $\tau_{kg}$
\[
\Lambda_{kg} = \frac{\pi^{-N_k M_g/2}\kappa^{1/2} \delta^{\gamma/2}}{(N_k M_g + \kappa)^{1/2}} \frac{\Gamma(\{N_k M_g + \gamma\}/2)}{\Gamma(\gamma/2)} \left(SS_{kg} + \kappa \xi^2 - \frac{(S_{kg}+\kappa \xi)^2}{N_k M_g + \kappa} + \delta \right)^{-(N_k M_g + \gamma)/2}.
\]
Here $S_{kg}=\sum\limits_{i:c_i=k} \sum\limits_{j:w_j=g} y_{ij}$ and $SS_{kg}=\sum\limits_{i:c_i=k} \sum\limits_{j:w_j=g} y_{ij}^2$.

\subsection{A greedy search strategy} \label{sec:greedy}

Based on the observations about the ICL in Section~\ref{sec:ex_ICL} it makes sense to try to find some strategy to optimise the exact ICL. There are in essence four unknowns in our problem $K,G,\bc,\bw$. From Section~\ref{sec:ex_ICL} the largest value of the exact ICL (assuming it can be computed), gives the clustering of the nodes into subsets which is most preferable. The idea used by~\citeasnoun{Come13} for the stochastic blockmodel was to optimise the ICL criterion using a greedy search over labels and the number of node clusters. We adopt a similar approach here. First define the ICL using any instance of numbers of clusters and labels
\[
\ICL(K,G,\bc,\bw) = \log \pi(\bc,\bw|K,G) + \sum_{k=1}^K \sum_{g=1}^G \log \Lambda_{kg}.
\]
This frames the exact ICL as an objective function in parameters $(K,G,\bc,\bw)$ which we wish to optimise. Of course, the values of $K$ and $G$ place constraints on the possible values of $\bc$ and $\bw$. The exact ICL is optimised iteratively by cycling through the following smaller optimisation techniques repeatedly until no further increases in it can be obtained. 

\subsubsection*{Update node labels in set $A$} 

\begin{figure}
\begin{center}
\fbox{
\begin{tikzpicture}
  \tikzstyle{sta} = [rectangle, rounded corners,draw,minimum width=20pt]

	\node[text width = 110pt] (prop) at (-1,5.2) {\bf{Compute ICL differences for change of label of node } $i$ \bf{ from } $k'$ \bf{ to } $l$, $\Delta_{k'\rightarrow l}$};
	
	\node[text width = 90pt] (propw) at (3.8,5.2) {\bf{Identify the best label update for node }$i$\bf{ based on ICL}};
	
	\node[text width = 80pt] (propr) at (7.5,5.2) {\bf{Move node }$i$\bf{ updating the label}};
	
	\draw[->] (propw) edge (propr);
	
	\node[text width = 110pt] (propq) at (-6,5.2) {\bf{Current labels and current ICL when node }$i$\bf{ in cluster }$k'$ };
	
	\draw [decorate,decoration={brace,amplitude=10pt,mirror},xshift=0pt,yshift=0pt]
(1.5,3.9) -- (1.5,6.2)node [black,midway] {};
	
	\draw[->] (propq) edge (prop);

	\node[sta] (orig1) at (-6,1) {$\mathbf{c},\mathbf{w}$ current labels}; 
	\node[sta] (mv1) at (-1,3) {$\mathbf{c}_{(1)}^{*},\mathbf{w}$ giving $\Delta_{k'\rightarrow 1}$}; 
  	\node[sta] (mv2) at (-1,1.5) {$\mathbf{c}_{(2)}^{*},\mathbf{w}$ giving $\Delta_{k' \rightarrow 2}$}; 
  	\node[sta] (mv3) at (-1,-1) {$\mathbf{c}_{(K)}^{*},\mathbf{w}$ giving $\Delta_{k' \rightarrow K}$}; 
  	
  	\node (mvd) at (-1,0.5) {$\vdots$};
  	\node (mvd) at (-1,0) {$\vdots$};
  	
  	\draw[->]   (orig1) edge (mv1);
  	\draw[->]  	(orig1) edge (mv2);
  	\draw[->] 	(orig1) edge (mv3);
  	
  	\draw [decorate,decoration={brace,amplitude=10pt,mirror},xshift=0pt,yshift=0pt]
(1.5,-1) -- (1.5,3)node [black,midway] {};

	\node[sta,text width = 90pt] (Expl) at (3.8,1) {Take cluster $l$ that gives max $\Delta_{k'\rightarrow l}$ and move node $i$ to this cluster (leave put if max $\Delta_{k'\rightarrow l}$ is 0)} ;
	

	\node[sta] (neworig1) at (8,1) {$\mathbf{c},\mathbf{w}$ updated labels}; 
	
	\draw[->] (Expl) edge (neworig1);

\end{tikzpicture}
}
\end{center}
\caption{Flowchart showing the process for updating the label of a node $i$ from set $A$. Here the $\mathbf{c}^*$ are subscripted with the candidate cluster for clarity. The top portion (boldface) shows gives  an overview of each stage of the process. The bottom portion outlines the computations carried out. } \label{fig:flowchart}
\end{figure}
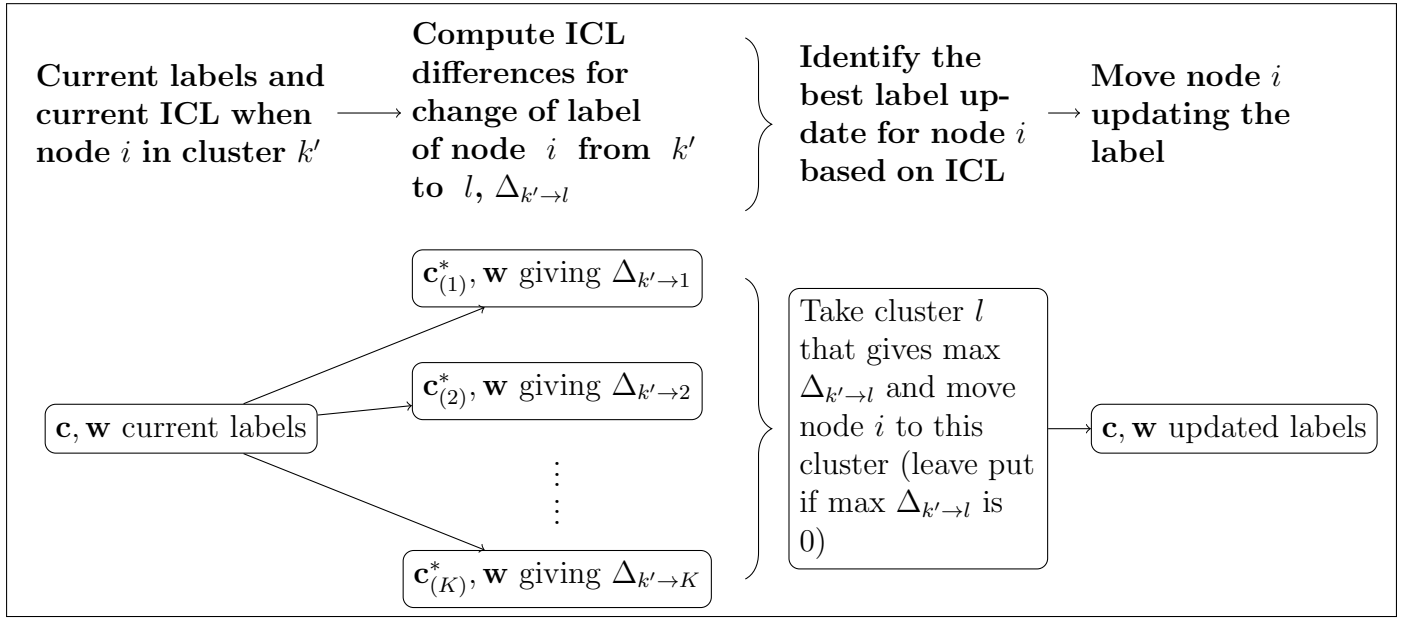

To update the node labels for set $A$, we firstly shuffle the order of nodes in the update. If node $i$ is currently in $A_{k'}$ (cluster $k'$), that is $c_i = k'$, compute the change in exact ICL when moving node $i$ to $A_l$ for all $l\ne k'$. First of all, suppose that currently $|A_{k'}|>1$, then the change in exact ICL is given by
\[
\Delta_{k' \rightarrow l} = \ICL(K,G,\bc^*,\bw) - \ICL(K,G,\bc,\bw)
\]
where $\bc^*$ is such that $c_r^* = c_r, r \ne i$ and $c_i^* = l$. Clearly $\Delta_{k' \rightarrow k'} = 0$. If all of the $\Delta_{k' \rightarrow l} < 0$, do not move node $i$ from subset $k'$. Otherwise, move $i$ to the node cluster $A_{l^*}$ where $l^* = \arg \max_l \Delta_{k' \rightarrow l}$.
This process is illustrated in a flow diagram in Figure~\ref{fig:flowchart} which summarizes the primary parts of the update.

If $|A_{k'}| = 1$, then our algorithm assumes that moving node $i$ from $A_{k'}$ causes that subset to vanish, so that $K$, the number of subsets is reduced by 1. This is similar to component absorption in the case of algorithms for finite mixture models. In this case, the computation of the change in exact ICL must be modified accordingly to 
\[
\Delta_{k' \rightarrow l} = \ICL(K-1,G,\bc^*,\bw) - \ICL(K,G,\bc,\bw).
\] 
This updating process is repeated for each node $i$ in $A$.

\subsubsection*{Update node labels in set $B$}

The node labels in set $B$ are updated in an analogous way to those in set $A$, so we omit a description for brevity. 

\subsubsection*{Algorithm initialization and termination}

The algorithm is initialized randomly, with a random assignment of nodes to the subsets $A_k , B_g, k=1\dots,K, g=1,\dots,G$. Initially we assume a large number of clusters of nodes. To draw analogy with the MCMC sampler described in Section~\ref{sec:lbm_est} we set $K=K_{\max}$ and $G=G_{\max}$ at the beginning. Nodes in sets $A$ and $B$ are then processed, and the algorithm terminates when no improvement (or further increase) in the exact ICL can be obtained. Clearly, the algorithm will cause convergence of the ICL to a local maximum, which may not necessarily be the global maximum. Thus, the greedy algorithm is run a number of times (this could be done in parallel), and the run given the highest exact ICL value used. As suggested by one reviewer, in practice, one could use another algorithm (for example the algorithm of~\citeasnoun{Rohe12} extended to the bipartite setting) to provide an initial clustering before invoking the greedy search algorithm. This may lead to faster convergence of the algorithm and quicker run times. 

\subsubsection*{Using a merging move at termination}

At termination of the algorithm, we can attempt to merge clusters if this increases the exact ICL further. Merging the clusters is more of a ``mass node'' move than the one node updates used in the greedy search. 
To merge two $A$ node clusters, we compute $\ICL(K-1,G,\tilde{\bc},\bw)$ where if attempting to merge clusters $k$ and $k'$, $\tilde{\bc}$ is such that all nodes with label $k'$ in $\bc$ become label $k$ nodes. This necessitates $G$ new block marginal likelihood calculations to get
\[
\ICL(K-1,G,\tilde{\bc},\bw) - \ICL(K,G,\bc,\bw) = \log\left\{\frac{\pi(\tilde{\bc},\bw|K-1,G)}{\pi(\bc,\bw|K,G)} \right\} + \sum_{g=1}^G \left(\log\tilde{\Lambda}_{kg} - \log (\Lambda_{kg}\Lambda_{k'g}) \right)
\]
where $\tilde{\Lambda}_{kg}$ is obtained by merging the sufficient statistics for blocks $(k,g)$ and $(k',g)$ and recomputing the marginal likelihood for these new statistics. If this difference is greater than 0 we merge the clusters $k$ and $k'$, otherwise no merge is performed. All pairwise merges are considered after the termination of the greedy exact ICL algorithm.

\subsection{Computational complexity of the greedy algorithm} \label{sec:complex}

Quantifying exactly the computational complexity of the greedy algorithm is not possible, as it will be problem dependent. However, we can analyse its various components and give some guidelines and comparison with the MCMC algorithm of~\citeasnoun{Wyse12}. Assume that the average cost of computing one of the $\Lambda_{kg}$ terms is $Q_{\Lambda}$ (this clearly depends on the model chosen for the linking attributes). To update node $i$ in cluster $A_{k'}$ we must compute $K-1$ terms of the form
\begin{eqnarray*}
\Delta_{k' \rightarrow l} & = & \ICL(K,G,\bc^*,\bw) - \ICL(K,G,\bc,\bw)\\
 & = & \left[\log \pi(\bc^*,\bw|K,G) + \sum_{k=1}^K \sum_{g=1}^G \log \Lambda_{kg}^* \right] - \left[\log \pi(\bc,\bw|K,G) + \sum_{k=1}^K \sum_{g=1}^G \log \Lambda_{kg} \right]
\end{eqnarray*} 
assuming that $|A_{k'}|>1$, where the $\Lambda_{kg}^*$ refer to the label vector $\bc^*$. One can see that $\Lambda_{kg}^* = \Lambda_{kg}$ for all but the affected subsets $k'$ and $l$, so that cancellation results in
\begin{eqnarray}
\Delta_{k' \rightarrow l} 
 & = & \left[\log \pi(\bc^*,\bw|K,G) + \sum_{g=1}^G \left(\log \Lambda_{k'g}^{(-i)} + \log \Lambda_{lg}^{(+i)} \right) \right] \nonumber  \\ 
 & & - \left[\log \pi(\bc,\bw|K,G) + \sum_{g=1}^G \left(\log \Lambda_{k'g} + \log \Lambda_{lg} \right) \right].\label{eq:change_icl}
\end{eqnarray}
Where now we have substituted the $\Lambda_{kg}^*$ for more meaningful and specific notation (``$+i$'' meaning add $i$ to the block, ``$-i$'' meaning take $i$ out of the block). For all of the linking attribute models mentioned in Section~\ref{sec:models}, one can store the sufficient statistics of each $(k,g)$ block to compute the $\Lambda_{kg}$, meaning that $\Lambda_{kg}^*$ can be computed by a slight modification of these statistics. Also we can store the $\Lambda_{kg}$ from the most recent state of the algorithm. The term $\log \pi(\bc^*,\bw|K,G) - \log \pi(\bc,\bw|K,G)$ is equal to 
\[
\log \left(\frac{\Gamma\{N_{k'}-1+\alpha_0\}\Gamma\{N_l + 1 + \alpha_0\} }{\Gamma\{N_{k'}+\alpha_0\}\Gamma\{N_l + \alpha_0\}}\right).
\]
We assume that the cost of computing this quantity is dominated by the computation of the $\Lambda_{kg}$. This is reasonable for all of the models in Section~\ref{sec:models}. This means that (\ref{eq:change_icl}) can be computed with $O(2 Q_{\Lambda} G)$ operations (for the case $|A_{k'}| = 1$ it can be shown that $O(Q_{\Lambda} G)$ operations are required for the calculation). So updating the label of node $i$ is $O(2Q_{\Lambda} K G)$. One can show that updating nodes in set $B$ has the same cost, giving $O(2 N M Q_{\Lambda} K G)$ overall per sweep of the greedy algorithm. If we compare to~\citeasnoun{Wyse12}, the total cost of the Gibbs update for nodes for $T_{\mathrm{it}}$ iterations is fixed at $O(2T_{\mathrm{it}}NMQ_{\Lambda} K G)$. Usually we would expect convergence of the greedy algorithm in much fewer than $T_{\mathrm{it}}$ iterations. Also there are other moves in~\citeasnoun{Wyse12}, which are costly but whose complexity is difficult to quantify exactly. In short, we expect our approach to be much more scalable than the MCMC based inference. Another point to note is that the performance in terms of mixing 
of the algorithm of~\citeasnoun{Wyse12} usually declines as the dimensions $N$ and/or $M$ increase, so that although we may be expending much more computing power, we may not necessarily be gaining much more information about the clustering. The MCMC algorithm also has limitations in size, in that to run it on larger datasets may not even be feasible from a time consideration.   

\subsection{Why the greedy search is both sensible and pragmatic} \label{sec:sensible}

Consider a Gibbs sampler for updating the labels of the nodes from set $A$ for example similar to that used in~\citeasnoun{Wyse12}. Using (\ref{eq:post}) the full conditional distribution of the new label for node $i$ with $c_i = k'$, is 
\[
\pi(c_i^* = l |\bc_{-i},\bw,Y,K,G) \propto \Gamma\{N_{k'}-1+\alpha_0\} \Gamma\{N_l+1+\alpha_0\}\prod_{g=1}^G \Lambda_{k'g}^{(-i)} \Lambda_{lg}^{(+i)}
\]
for $l \ne k$ where $\bc_{-i}$ denotes $\bc$ without the $i^{\mathrm{th}}$ entry and $\Lambda_{k'g}^{(-i)}$ denotes the quantity computed by removing the statistics associated with node $i$. Also 
\[
\pi(c_i^* = k' |\bc_{-i},\bw,Y,K,G) \propto \Gamma\{N_{k'}+\alpha_0\} \Gamma\{N_l+\alpha_0\}\prod_{g=1}^G \Lambda_{k'g} \Lambda_{lg}
\]
reflects no change in label. Notice that 
\[
\frac{\pi(c_i^* = l |\bc_{-i},\bw,Y,K,G)}{\pi(c_i^* = k' |\bc_{-i},\bw,Y,K,G)} \propto \frac{\Gamma\{N_{k'}-1+\alpha_0\} \Gamma\{N_l+1+\alpha_0\}}{ \Gamma\{N_{k'}+\alpha_0\} \Gamma\{N_l+\alpha_0\}} \frac{\prod_{g=1}^G \Lambda_{k'g}^{(-i)} \Lambda_{lg}^{(+i)}}{\prod_{g=1}^G \Lambda_{k'g} \Lambda_{lg}} = \exp\{\Delta_{k'\rightarrow l}\}
\]
from (\ref{eq:change_icl}). Using this relation, it is straightforward to show that
\begin{equation}
\pi(c_i^* = l |\bc_{-i},\bw,Y,K,G) = \frac{\exp\{\Delta_{k'\rightarrow l}\}}{\sum_{k=1}^K \exp\{\Delta_{k'\rightarrow k}\}}. \label{eq:fullcond}
\end{equation}
This implies that changing the label of node $i$ to the label that maximizes the change in exact ICL (or alternatively not changing the node label if none of the changes in exact ICL are positive) corresponds to maximizing the full conditional distribution of the label in the Gibbs sampler. So instead of stochastically sampling a label from the full conditional, we deterministically choose the label that maximizes it in our greedy exact ICL search. This is similar to the iterated conditional modes algorithm of~\citeasnoun{Besag86}. For this reason we argue that using the exact ICL as an objective function is sensible. The pragmatic nature of the algorithm comes from the reasoning that if there is a strong cohesion to subsets (strong clustering and good separation of node subsets), then the label that will be sampled from the full conditional will usually be the one which corresponds to the largest value of the change in exact ICL. If our objective is to get the optimal partitioning of the node sets, then this 
appears an appealing approach. 

\subsection{Computational savings for the greedy ICL algorithm}\label{sec:sec4}

There are two main ways we can reduce the computational complexity of the greedy algorithm discussed in Section~\ref{sec:complex}. The first of these concerns the correspondence between maximizing exact ICL and the full conditional distribution of labels as shown in Section~\ref{sec:sensible}. The second is by exploiting the fact that often observed bipartite networks are sparse.

\subsubsection{Greedy search pruning}\label{sec:pruning}

One might expect that after a few sweeps of the greedy algorithm, some degree of degeneracy will appear in the full conditional distributions of the labels. In such situations one will observe that $\pi(c_i^* = l |\bc_{-i},\bw,Y,K,G)$ will be quite small compared with the full conditional for some other label, say $\pi(c_i^* = k' |\bc_{-i},\bw,Y,K,G)$. Suppose that currently $c_i=k$. Compute $\Delta_{k\rightarrow l}$ for all $l \ne k$, where $\Delta_{k\rightarrow k} = 0$. Let $k' = \arg\max_{r} \Delta_{k \rightarrow r}$. Consider labels $l$ which are very unlikely for node $i$ compared to $k'$
\[
\frac{\pi(c_i^* = l |\bc_{-i},\bw,Y,K,G)}{ \pi(c_i^* = l |\bc_{-i},\bw,Y,K,G)+\pi(c_i^* = k' |\bc_{-i},\bw,Y,K,G)} < \varepsilon
\]
for some small number $\varepsilon$. Using (\ref{eq:fullcond}) this may be written
\[
\frac{\exp\{\Delta_{k\rightarrow l}\}}{\exp\{\Delta_{k\rightarrow l}\} + \exp\{\Delta_{k\rightarrow k'}\}} < \varepsilon.
\]
This can be rearranged to give
\begin{equation}
\Delta_{k \rightarrow k'} - \Delta_{k \rightarrow l} > T(\varepsilon) \label{eq:thresh}
\end{equation}
where $T(\varepsilon)$ is a threshold depending on the value of $\varepsilon$. To reduce our search space in future iterations of the greedy search, we ignore computing the exact ICL for label changes to $l$ when (\ref{eq:thresh}) is satisfied. In this paper we take $T(\varepsilon) = 150$. This appeared to give satisfactory results without excluding important parts of the search space and introducing error. It can be seen that removal of these poor configurations from the search space will reduce the $O(4Q_{\Lambda}KG)$ cost of optimizing the labels at each step. The procedure has been described here for nodes from set $A$ but applies analogously to those in set $B$.

\subsubsection{Sparse representations}\label{sec:sparse}

When dealing with observed bipartite networks one often encounters much sparsity. By sparsity we mean many observed non-ties. For this reason, it can be useful to use a sparse representation of the adjacency matrix $Y$, denoted $Y^{\,\mathrm{s}}$, and use this to perform the exact ICL calculations outlined earlier. Only the non-zero valued ties of $Y$ are stored in $Y^{\,\mathrm{s}}$. For example, suppose we would like to compute the ICL difference given in (\ref{eq:change_icl}) for $Y^{\,\mathrm{s}}$ stored in sparse form. We denote a non-tie generically by ``0''. Suppose there are $m_i$ non-zero entries in the row corresponding to node $i$ (from $A$) of $Y^{\,\mathrm{s}}$ in columns $j_1^i,\dots,j_{m_i}^i$. Let the indexes of the non-zero columns  which correspond to nodes from $B$ and are members of $B_g$ be stored in the set $\mathcal{J}_g^i = \{ l : w_{j_l} = g\}$ and denote the nodes that have corresponding zero columns and which are members of $B_g$ by $\tilde{\mathcal{J}}_g^i = \{l:w_l = g\}
\backslash \mathcal{J}_g^i$. For each $g$ compute
\[
\upsilon_{g}^{i\,:\,\mathrm{nz}} = \sum_{l=1}^{m_i} \mathrm{I}(w_{j_l^i} = g)
\]
which gives the number of non-zero entries in $B_g$ which in row $i$ of the adjacency matrix, corresponding to node $i$. Then the number of zero entries in columns indexed by $B_g$ for row $i$ is $\upsilon_g^{i\,:\,\mathrm{z}} = \upsilon_g - \upsilon_{g}^{i\,:\,\mathrm{nz}}$. Then
\begin{eqnarray}
\Lambda_{kg} & = & \int \pi(\boldtheta_{kg}) \prod_{i:c_i = k} \prod_{j:w_j = g} p(y_{ij}|\boldtheta_{kg})\,\diff \boldtheta_{kg}\nonumber\\
& = & \int \pi(\boldtheta_{kg}) \prod_{i:c_i = k} \left[ \prod_{j \in \mathcal{J}_g^i} p(y_{ij}|\boldtheta_{kg}) \prod_{j\in \tilde{\mathcal{J}}_g^i} p(y_{ij}|\boldtheta_{kg}) \right]\,\diff \boldtheta_{kg} \nonumber\\
& = & \int \pi(\boldtheta_{kg}) \prod_{i:c_i = k} \left[ p(\mbox{``0''}|\boldtheta_{kg})^{\upsilon_{g}^{i\,:\,\mathrm{z}}}\prod_{j \in \mathcal{J}_g^i} p(y_{ij}^{\,\mathrm{s}}|\boldtheta_{kg}) \right]\,\diff \boldtheta_{kg}.
\end{eqnarray}
Thus by storing the number of zeros, and carrying out the calculation for these terms collectively, we can save many evaluations. In the experiments below, we found that exploiting the sparsity of some problems gave a considerable speed up of the algorithm. 

\subsection{Algorithm types}

The greedy search algorithm can be split into four types outlined in Table~\ref{tab:algs}, algorithms A2 and A3 using the pruning ideas of Section~\ref{sec:pruning}, and A1 and A3 using the sparse representations of Section~\ref{sec:sparse}. Observing the algorithm types, A0 is the naive implementation and we would expect this to be the slowest. On the other end of the scale, we would expect A3 to be fastest. Between A1 and A2, we would expect A1 to be faster than A2 when the matrix is quite sparse, and A2 to be faster than A1 when there is strong separation of nodes into subsets (strong cohesion) $\mbox{i.e.}$ very strong blocking. For small to moderate sizes of adjacency matrices, differences in run times may not be apparent, but for larger and sparser adjacency matrices we would expect to notice differences. When updating labels of $A$ nodes or $B$ nodes, these are processed in a random order. This introduces some amount of stochasticity into the ultimate destination of the greedy search. This will be 
investigated in due course in the examples below. For the examples we take $\alpha_0$ and $\beta_0$ both equal to 1 throughout. For the pruning algorithms we allow five full sweeps of the data before pruning begins. 

\begin{table}
\begin{center}
\begin{tabular}{ccc}
\hline \hline
Algorithm & Pruning & Sparse form\\
\hline
A0 & No  & No \\
A1 & No & Yes \\
A2 & Yes & No \\
A3 & Yes & Yes\\
\hline \hline
\end{tabular}
\end{center}
\caption{Different types of greedy search algorithm.}\label{tab:algs}
\end{table}

\section{Simulation study}\label{sec:simstud}

To investigate the performance of the greedy search approaches, we compared them with the MCMC approach of~\citeasnoun{Wyse12} on simulated data.

\subsection{Study set-up}
The data was generated from the model outlined in Section~\ref{sec:sec2}, with $K=5$ and $G=5$. The label distributions were chosen as $\omega_k = 1/5 , k=1,\dots,5$ and $\rho_g = 1/5, g=1,\dots,5$. The node sets were chosen to be of size $N=100$ and $M=100$, giving a $100 \times 100$ adjacency matrix. Using the absent/present linking attribute outlined in Section~\ref{sec:models}, the ties were generated so as to give different levels of overlap in the blocks. There are five blocks of varying clarity in the matrix following
\[
\theta_{kg} = \left\{ \begin{array}{cl}
1-q & \mbox{ if }k=g\\
q & \mbox{ otherwise}
\end{array} \right.
\]
where we let $q$ vary from 0.0125 to 0.5 in steps of 0.0125, giving 40 values of $q$ in total. As $q$ gets closer to 0.5 the clustering task becomes more difficult, as the probability of a tie becomes constant across all blocks. Twenty datasets were generated at each value of $q$. 

\subsection{Comparing clusterings}

As a means of comparing different clusterings of the data, we extend the normalized mutual information measure introduced in~\citeasnoun{Vinh10} and used by, for example~\citeasnoun{Come13}.~\citeasnoun{Vinh10} provide extensive and rigorous justification for using this such a measure for clustering comparison. We define a combined measure for the two node sets (rows and columns of the adjacency) to account for the fact that we are effectively doing two clustering tasks simultaneously. Consider first node set $A$. Denote the true clustering (from the simulation) by the labels $\bc^{\mathrm{t}}$ and the estimated one (from the algorithm in question) by $\bc^{\mathrm{e}}$. The mutual information between the two clusterings is 
\[
\mathcal{I}_A(\bc^{\mathrm{e}},\bc^{\mathrm{t}}) = \sum_{k,l}^K P_{kl} \log\left(\frac{P_{kl}}{P_k^{\mathrm{e}} P_l^{\mathrm{t}}} \right)
\]
where 
\[
P_{kl} =\frac{1}{N} \sum_{i,j}^N \mathrm{I}(c_i^{\mathrm{e}} = k,c_j^{\mathrm{t}}=l),\, P_k^{\mathrm{e}} = \frac{1}{N} \sum_{i=1}^N \mathrm{I}(c_i^{\mathrm{e}}= k),  \, P_l^{\mathrm{t}} = \frac{1}{N} \sum_{i=1}^{N} \mathrm{I}(c_i^{\mathrm{t}} = l).
\] 
The mutual information measures how much is learned about the true clustering if the estimated one is known, and {\it vice versa}. We normalize this quantity when the clusterings can have a different number of clusters (as in our case)
\[
\mathcal{NI}_A(\bc^{\mathrm{e}},\bc^{\mathrm{t}})  = \frac{\mathcal{I}(\bc^{\mathrm{e}},\bc^{\mathrm{t}})}{\max(\mathcal{H}(\bc^{\mathrm{e}}),\mathcal{H}(\bc^{\mathrm{t}}))}
\]
with $\mathcal{H}_A(\bc) = - \sum_{k=1}^K P_k \log P_k$ and $P_k = \frac{1}{N} \sum_{i=1}^N \mathrm{I}(c_i=k)$.

We compute the same quantity analogously for node set $B$, and add the two together to give an overall measure of mutual information between estimated and true clusterings as 
\[
\mathcal{NI}_A(\bc^{\mathrm{e}},\bc^{\mathrm{t}}) + \mathcal{NI}_B(\bw^{\mathrm{e}},\bw^{\mathrm{t}})
\]
which has a maximum value of 2 with high agreement and minimum 0 with little or no agreement. 

\subsection{Different approaches}

\begin{figure}
\begin{center}
\includegraphics[width=100mm]{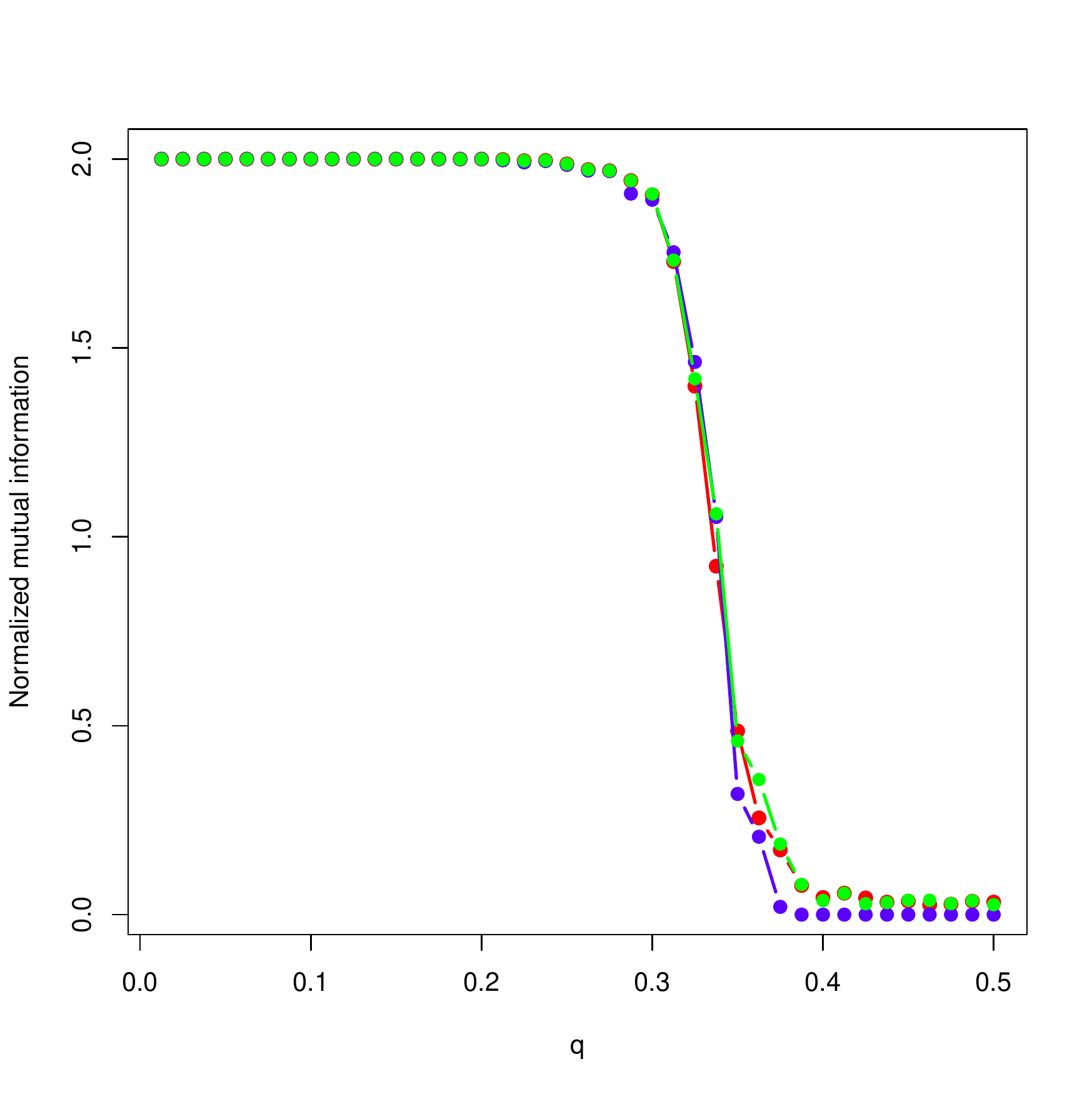}
\caption{Comparison of the normalized mutual information for the three algorithms. MCMC is shown in blue, greedy algorithm in red and the pruned greedy algorithm in green. }\label{fig:comparision}
\end{center}
\end{figure}

The three different approaches taken were 
\begin{itemize}
\item a run of the collapsed MCMC algorithm of~\citeasnoun{Wyse12} of 25,000 iterations taking the first 5,000 as a burn in
\item the basic greedy search algorithm A0 
\item the pruned greedy search algorithm A2.
\end{itemize}
The reason the sparse forms (A1 and A3) of the greedy search algorithms are not used here is the dimension of the adjacency matrix; a $100\times 100$ matrix runs quickly enough using algorithms A0 and A2 that exploiting any possibly sparsity is not of huge benefit here. 
For the MCMC algorithm, the estimated clustering was computed by first finding the MAP estimates of $K$ and $G$ from the approximated posterior, and then finding the MAP of the labels conditioning on these values.  It should be noted here that the MCMC takes considerably longer to run than either of the greedy algorithms, which are about 2,000 times faster in this instance. The pruned algorithm is usually about 10 times faster than the basic greedy algorithm. Figure~\ref{fig:comparision} shows the average value of the normalized mutual information over the twenty datasets for each value of $q$ for each of the algorithms. The MCMC algorithm is shown in blue, the greedy algorithm in red and the pruned greedy algorithm in green. It can be seen that as the value of $q$ approaches 0.5, the greedy algorithms slightly outperform the MCMC algorithm, with the MCMC outperforming the greedy algorithm around $q=0.35$. However, the observation most of note here is that the greedy algorithm gives essentially the same 
information as the longer MCMC run in a very small fraction of the time. 

\section{Examples}\label{sec:sec5}

\subsection{Congressional voting data}

First of all the congressional voting data was analysed by treating abstains as nay's. The data matrix is $435\times 16$. Here the node sets to be grouped are the congressmen (435) and the issues which they vote on (16). Ten instances of the algorithms in Table~\ref{tab:algs} were run, each with two restarts and the maximum ICL recorded from the ten runs. If restarting the algorithm affects the results, one would expect this to me more pronounced in the pruned versions A2 and A3, since restarting here corresponds to resetting the search for each row/column to the set of all possible clusters as opposed to the pruned set. 

To allow for direct comparison of the four algorithms, the random number generator was seeded the same- this also allows a direct time comparison for the same search progression as well as the maximum exact ICL reached. Table~\ref{ref:cong_results} shows the average runtime and maximum exact ICL reached for each algorithm over the ten runs. It can be seen that the sparse versions of the algorithm give a reduction in runtime with no noticeable reduction due to pruning.

\begin{table}
\begin{center}
\begin{tabular}{cccc}
\hline \hline
Algorithm & maximum ICL & average time (sec) & $(K,G)$\\
\hline
A0 & -3543.062  & 0.62 & (6,12) \\
A1 & -3543.062  & 0.56 & (6,12)\\
A2 & -3543.062  & 0.62 & (6,12)\\
A3 & -3543.062  & 0.56 & (6,12)\\
\hline \hline
\end{tabular}
\end{center}
\caption{Results for ten runs of each of the four algorithms on Congressional voting data with same random seed.} \label{ref:cong_results}
\end{table}

\begin{figure}
\includegraphics[width=40mm]{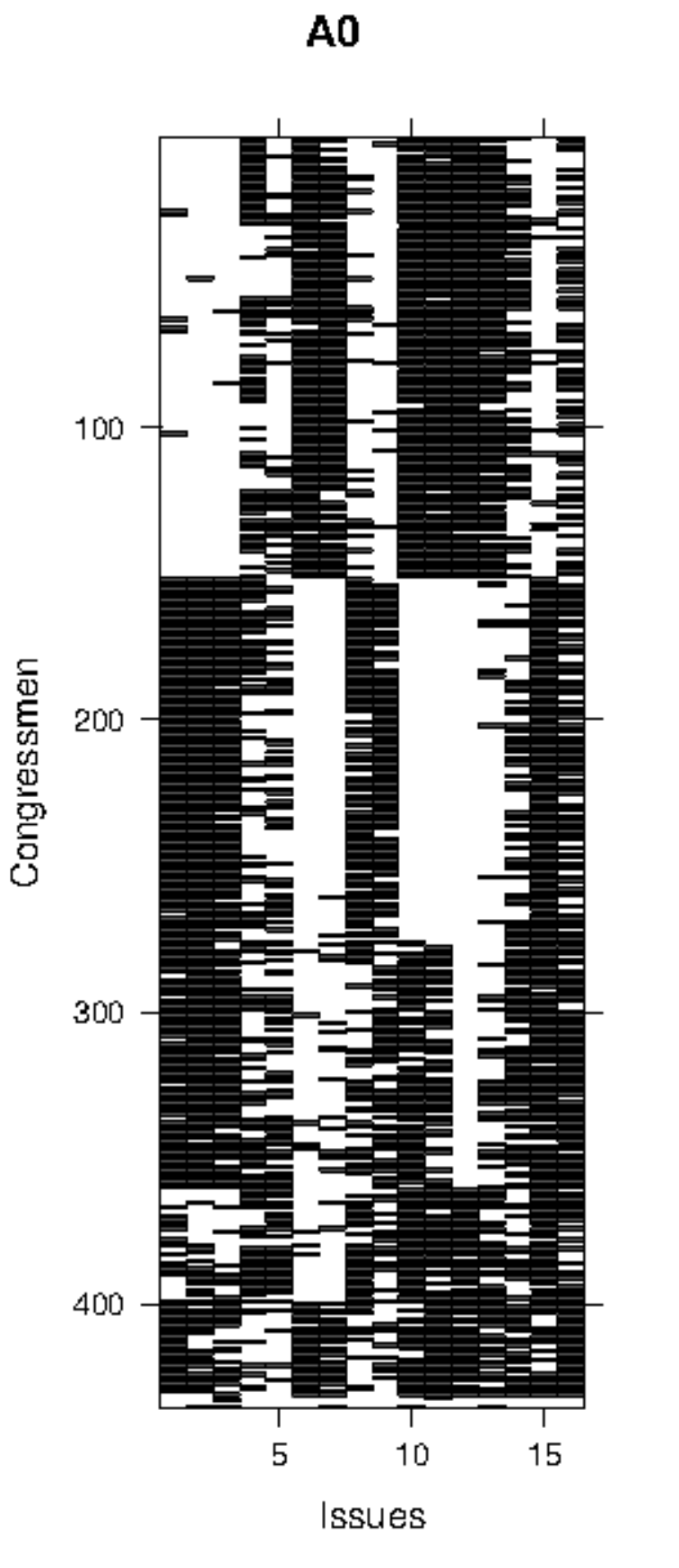} 
\includegraphics[width=40mm]{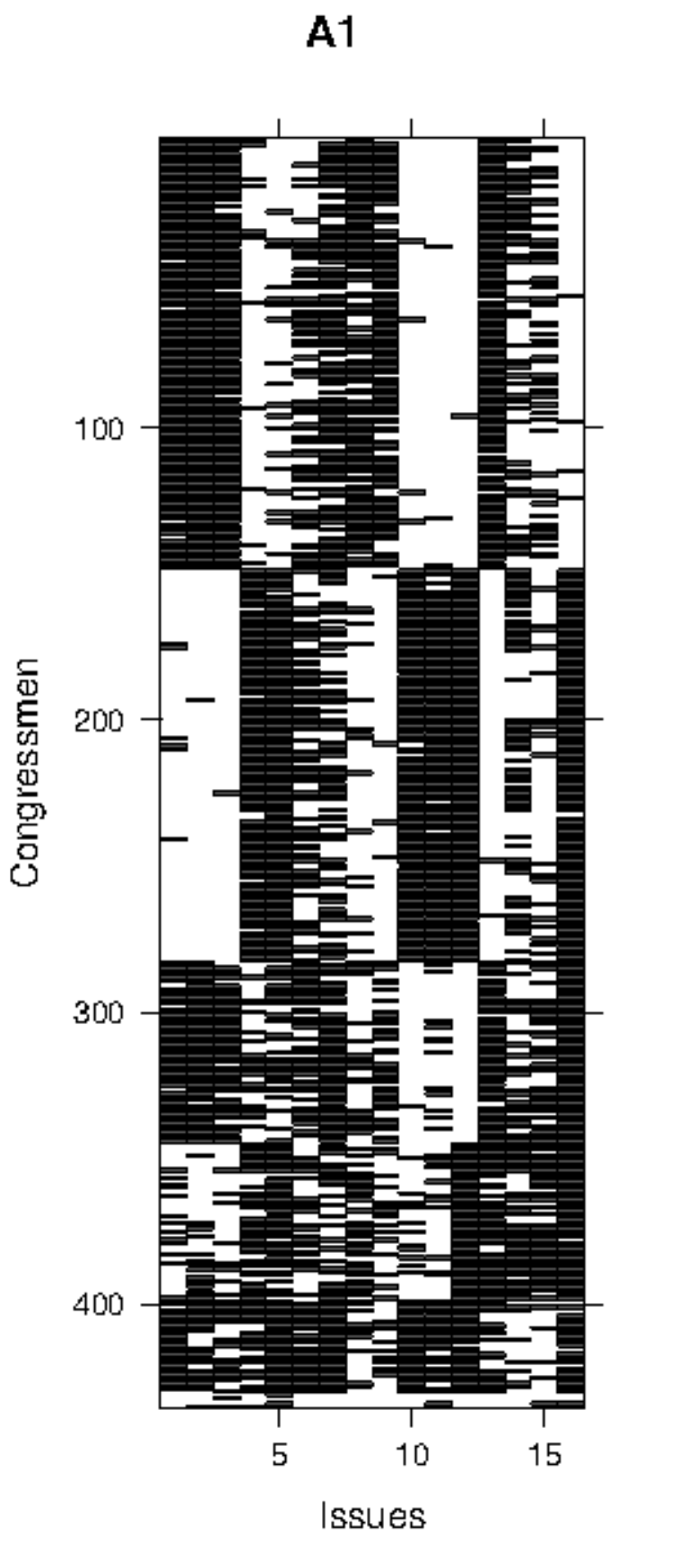} 
\includegraphics[width=40mm]{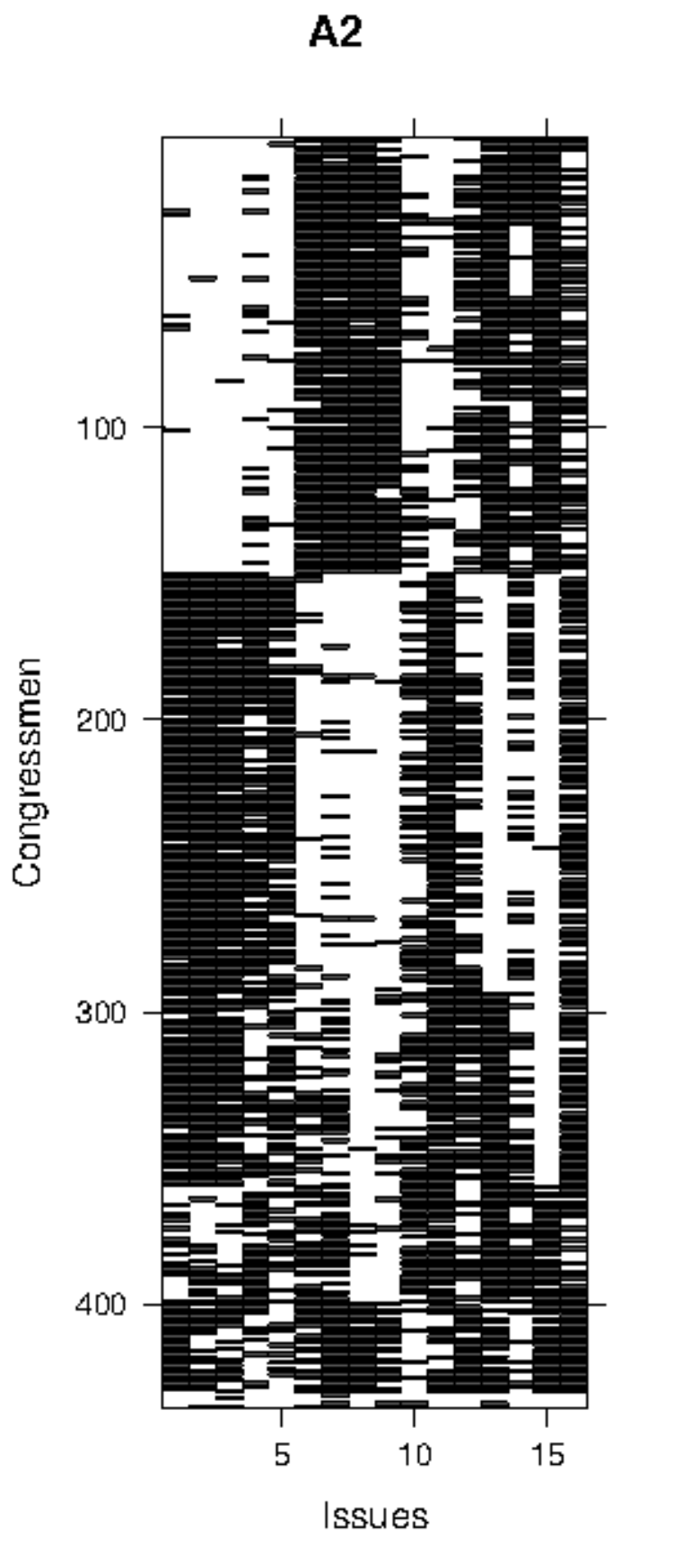} 
\includegraphics[width=40mm]{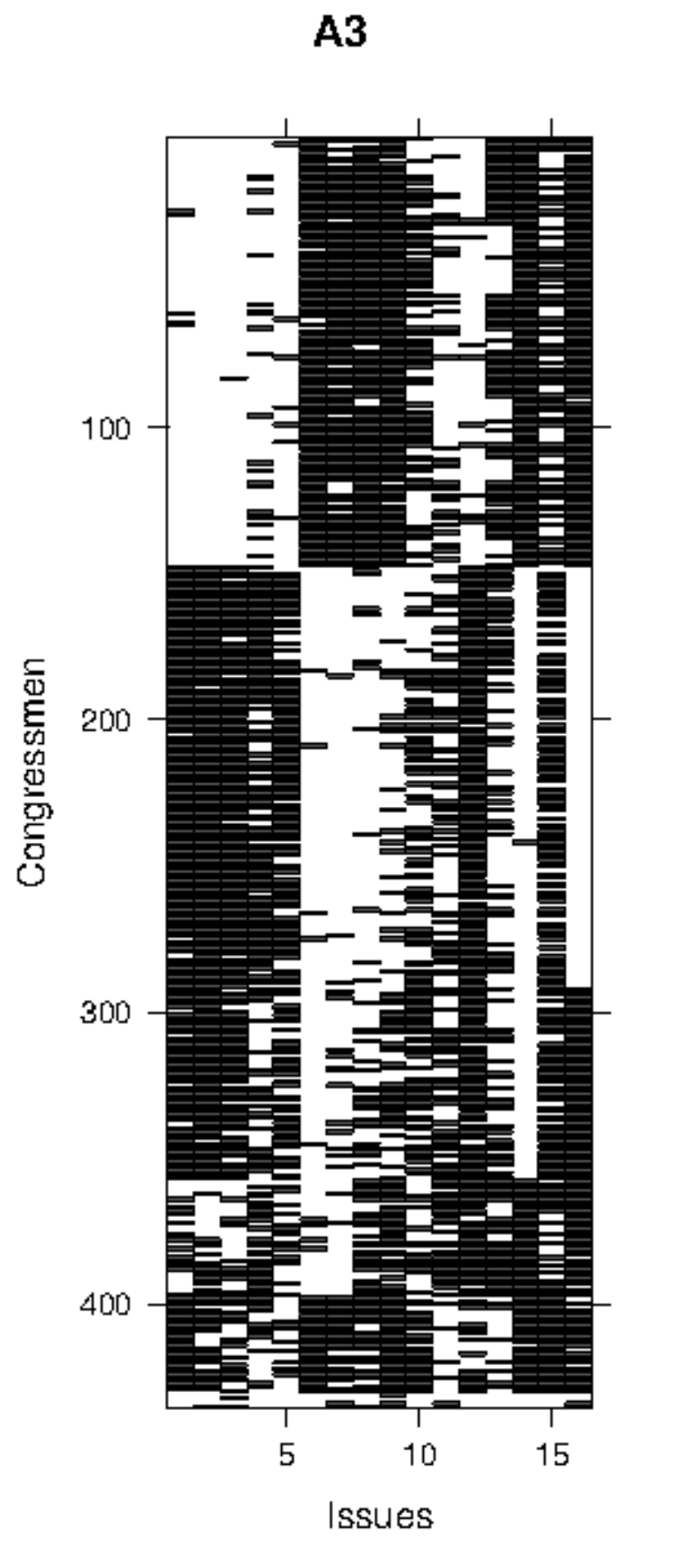} 
\caption{Re-ordered data matrix of maximum ICL configuration output over the ten runs of each algorithm.} \label{fig:cong_res}
\end{figure}

The experiment was repeated, this time with different random seeds for each algorithm. Figure~\ref{fig:cong_res} shows the re-ordered data matrices for the four highest ICL over ten runs of the four algorithms. Here the effect of the stochasticity on the outcome of the greedy search is noted. The four algorithms all converge to different numbers of clusters, with different resulting maximum exact ICL. Visually, all four algorithms appear to give a sensible result despite the fact that there are some differences. This highlights the possibility of many local maxima in general applications and the need to run the algorithm a number of times to give it the best chance to find a close to global maximum. 

\begin{table}
\begin{center}
\begin{tabular}{cccc}
\hline \hline
Algorithm & maximum ICL & average time (sec) & $(K,G)$\\
\hline
A0 & -3546.968  & 0.63 & (6,11) \\
A1 & -3543.812  & 0.58 & (6,10)\\
A2 & -3537.503  & 0.65 & (6,11)\\
A3 & -3537.550  & 0.64 & (6,11)\\
\hline \hline
\end{tabular}
\end{center}
\caption{Results for ten runs of each of the four algorithms on Congressional voting data with different random seed.} \label{ref:cong_results_diff}
\end{table}

\begin{figure}
\includegraphics[width=40mm]{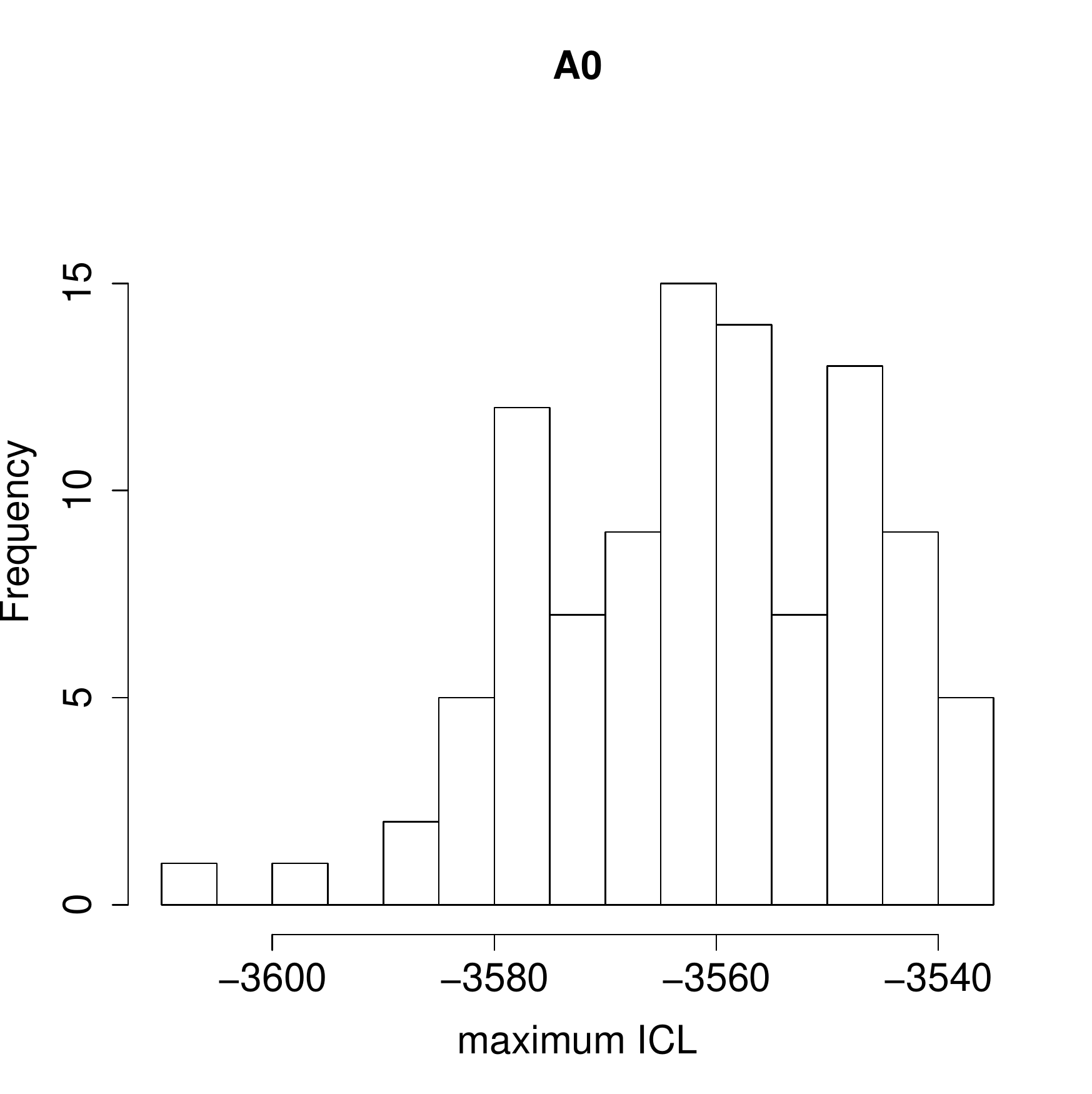}
\includegraphics[width=40mm]{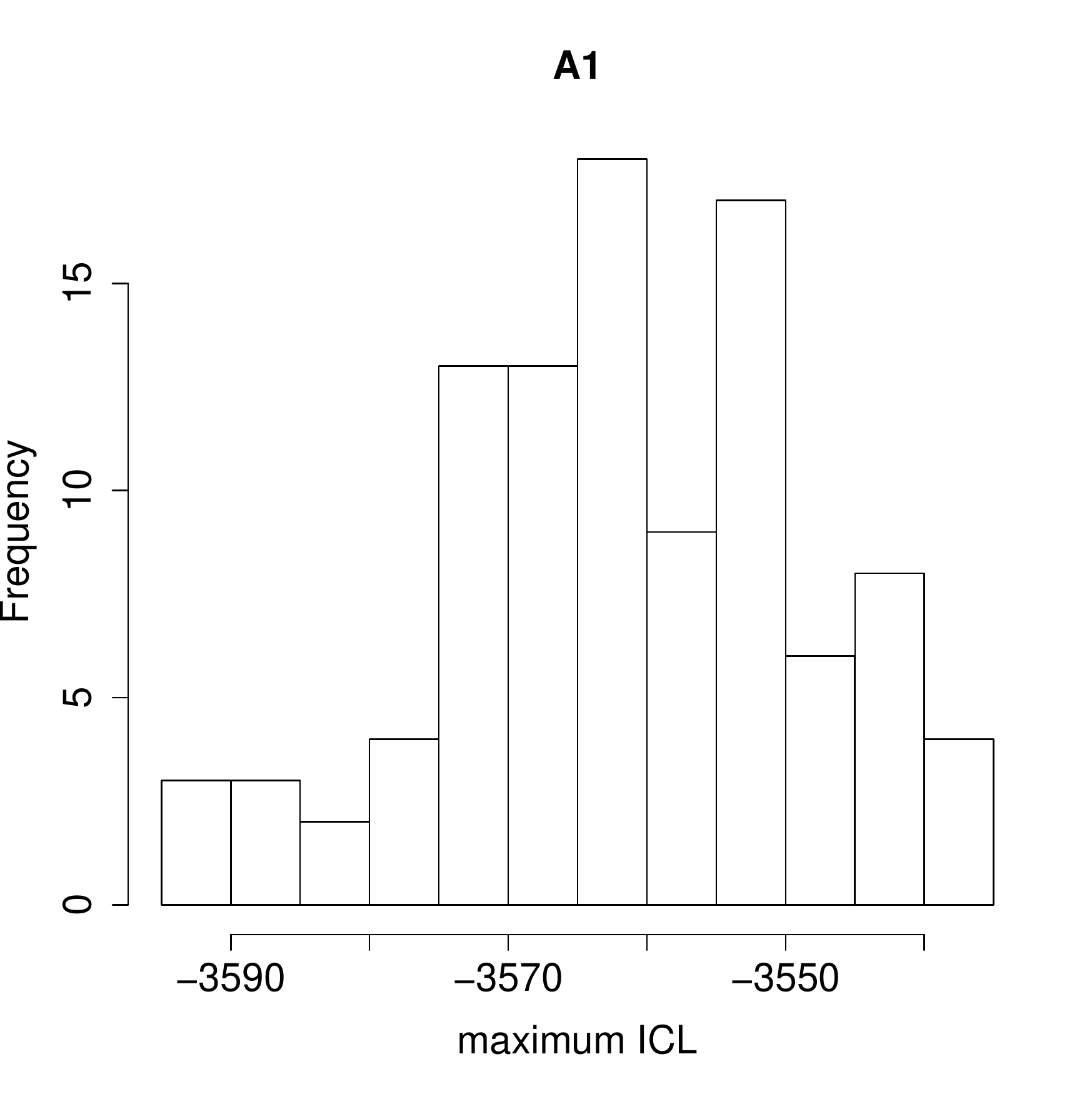}
\includegraphics[width=40mm]{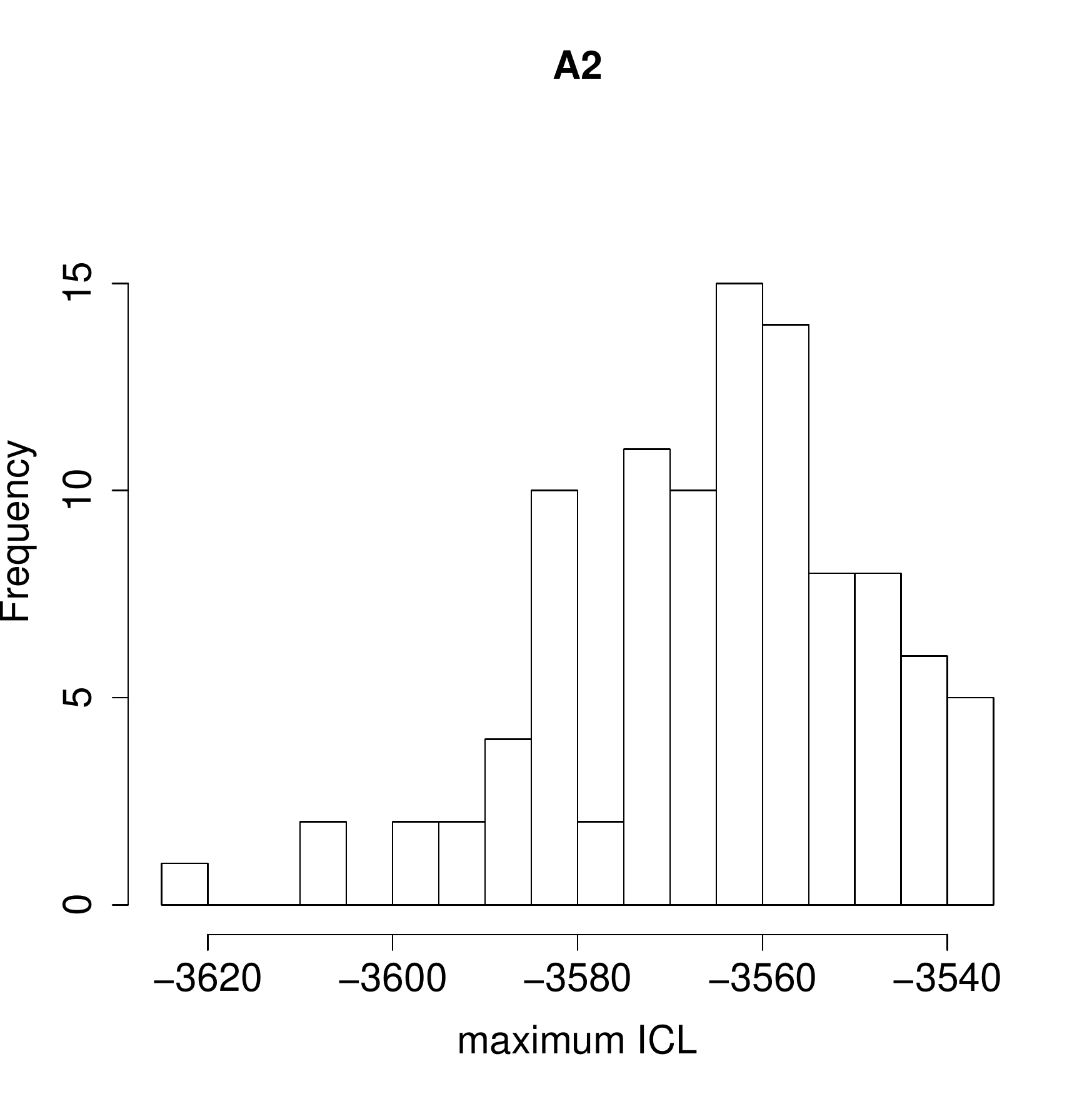}
\includegraphics[width=40mm]{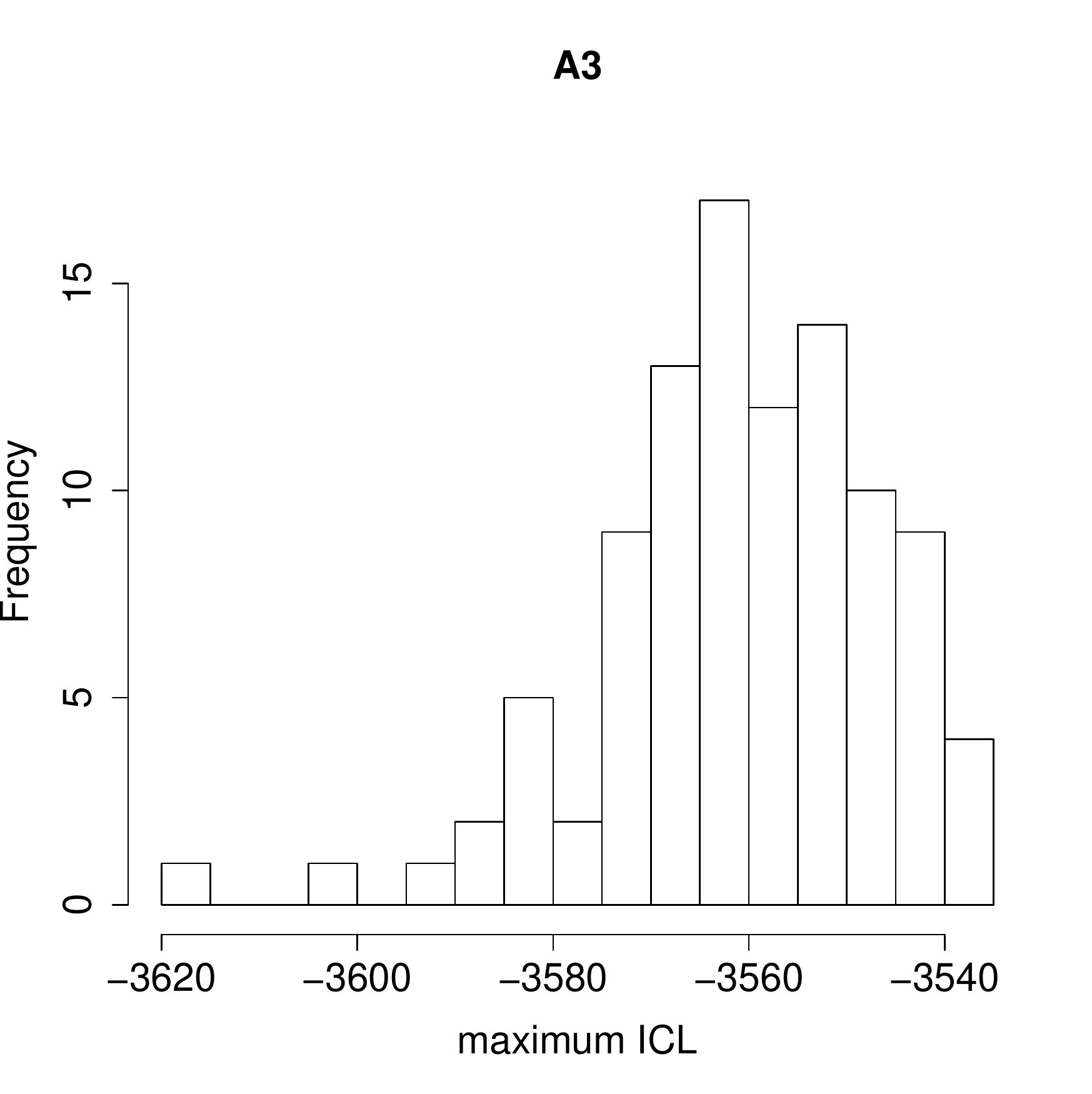}
\caption{Histograms of maximum ICL for 100 runs of each of algorithms A0-A4.} \label{fig:cong_hist}
\end{figure}

To take a closer look at the effect of starting value and the randomness introduced into the greedy search by processing nodes in a random fashion when updating the labels, we took 100 runs of each algorithm. Histograms of the maximum ICL obtained are shown in Figure~\ref{fig:cong_hist}. From this it can be seen that all four algorithms most often get to a maximum exact ICL around $-3560.00$ and less frequently obtaining values larger than this. As mentioned in Section~\ref{sec:greedy} and suggested by one reviewer, the optimal ICL could potentially be improved by providing a good starting value for the algorithm. In the absence of such an initialization, we'd recommend running the algorithm a number of times (10 or more) and taking the output of the run that gives the maximum ICL over these. This is similar to the general approach when the EM algorithm is fitted to finite mixture models; the run which results in the maximum value of the likelihood is taken as the output run. 

As a rough comparison to the MCMC based algorithm of~\citeasnoun{Wyse12}, their analysis took about thirty minutes on this network, gleaning much the same information that we have obtained here in a fraction of the time. The MCMC analysis here was coded in parallel with the implementation for computing the exact ICL and is more efficient than the original implementation in~\citeasnoun{Wyse12}, hence the difference in reported run time. The mixing of their algorithm, even for this moderately sized data was was slow for the congressman dimension with only about 1.5\% of proposed cluster additions/deletions being accepted. In terms of scalability, MCMC is not an option when dealing with adjacency matrices of any large size. The fact that both the MCMC approach and the exact ICL approach are based on almost identical posteriors indicates that the greedy search is the only viable option of the two for larger applications.

\subsection{Movie-Lens 100k}

\begin{figure}
\begin{center}
\includegraphics[width=50mm]{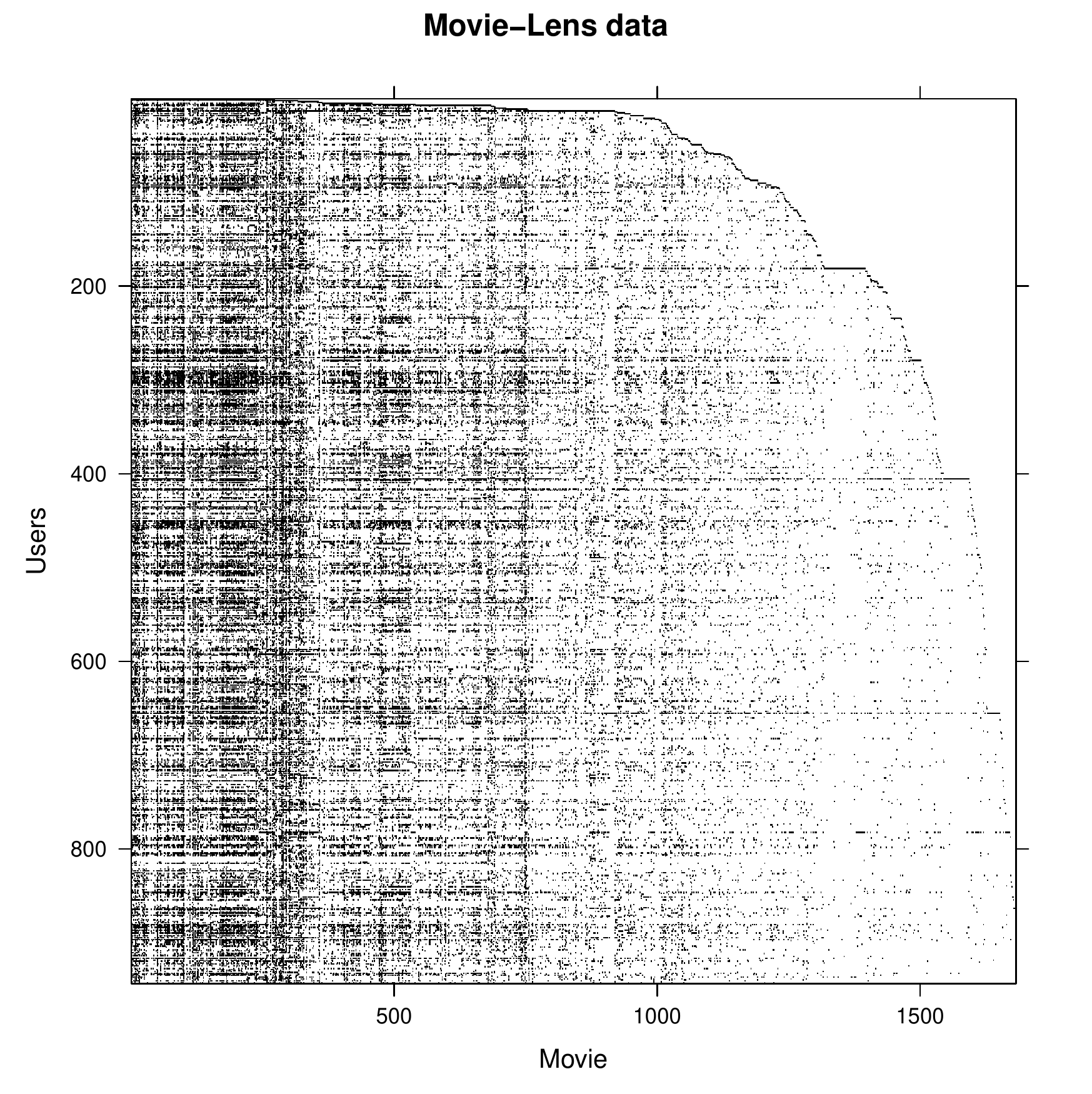}
\includegraphics[width=50mm]{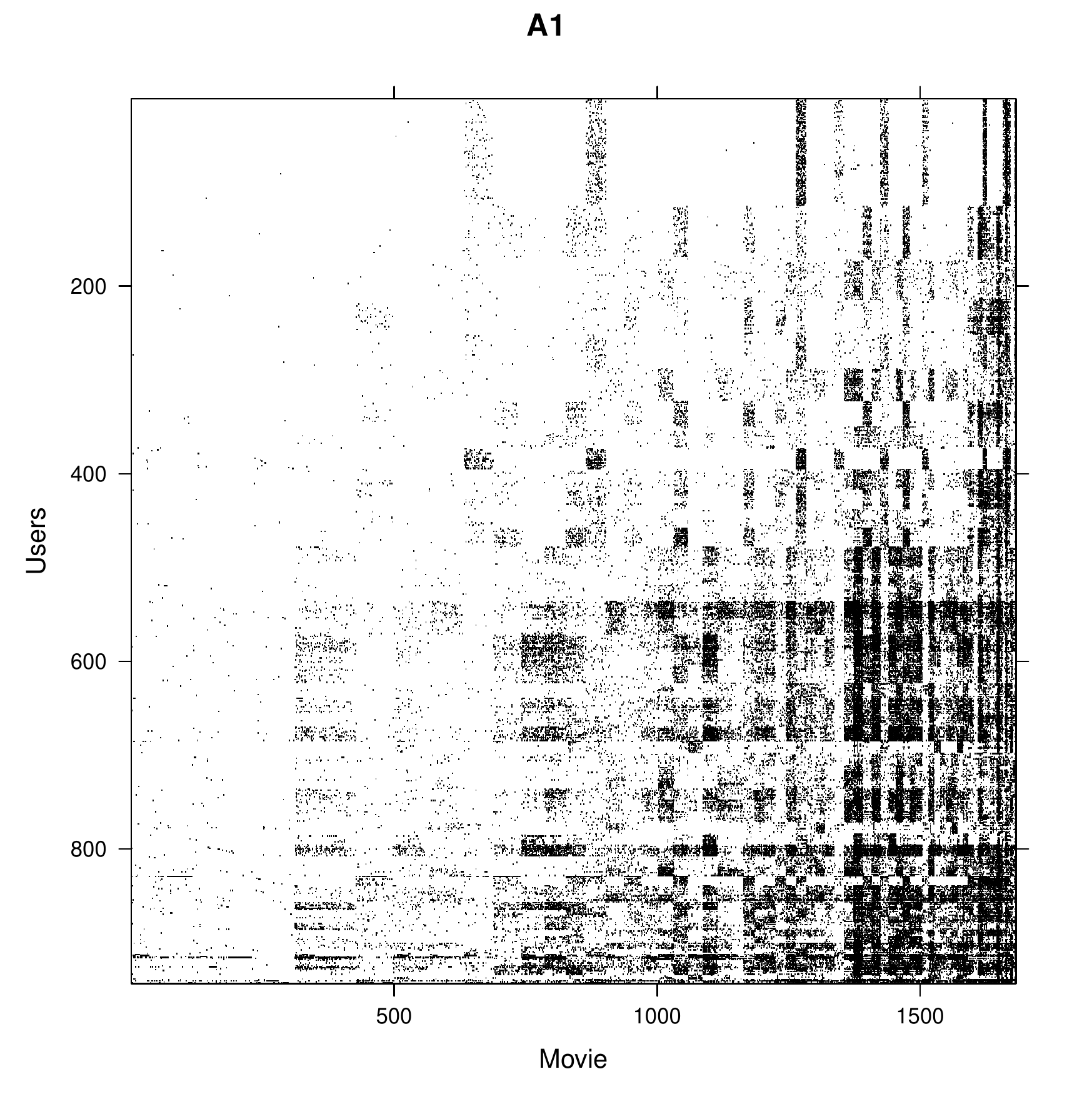}
\includegraphics[width=50mm]{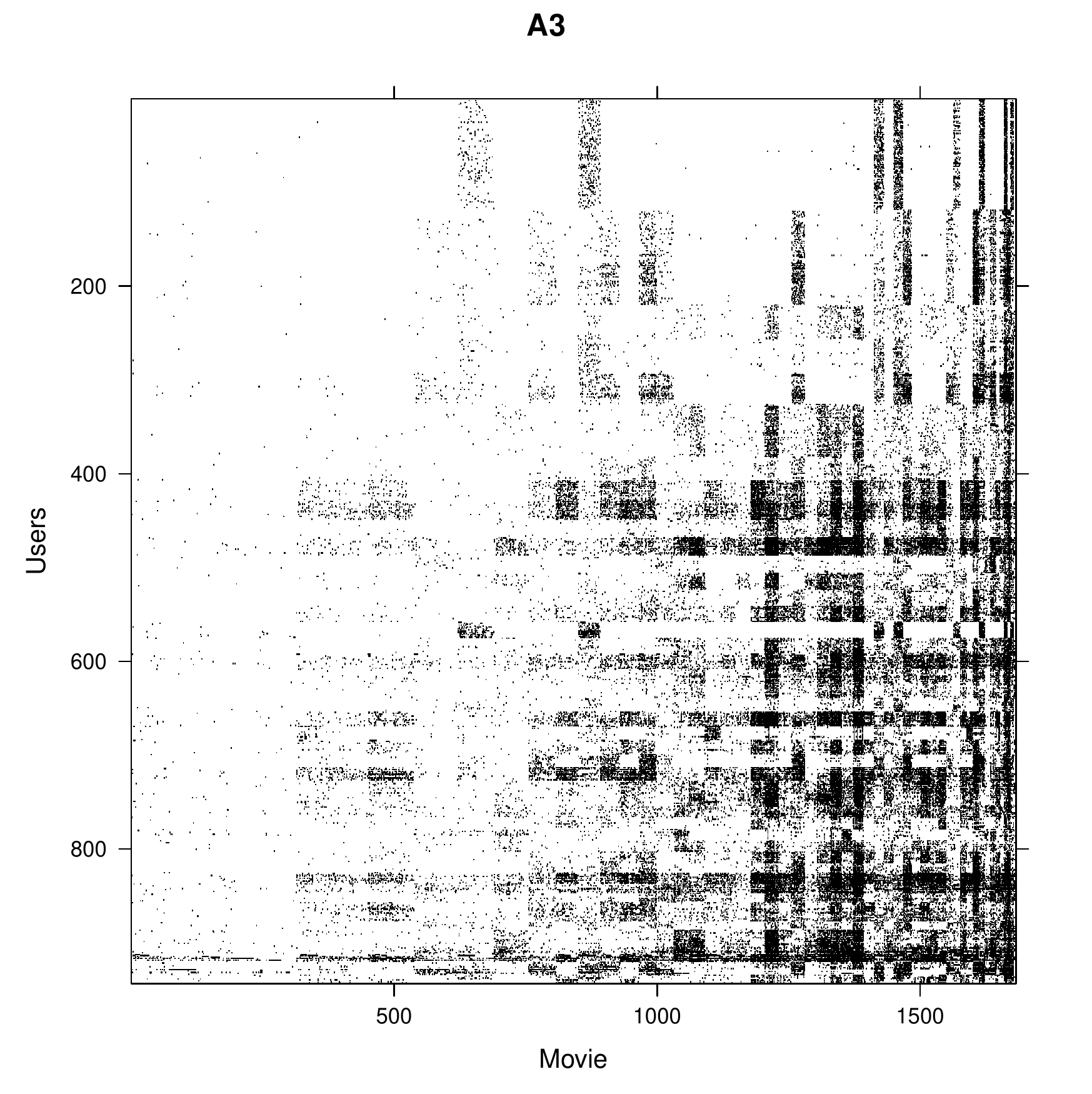}
\end{center}
\caption{The Movie-lens data with 100,000 user ratings and results from the A1 and A3 algorithm.} \label{fig:ml100k}
\end{figure}

The Movie-Lens dataset gives 100,000 ratings of movies by users. In total, there are 943 users and 1,682 movies. Many users do not rate particular movies so the network is quite sparse (93\%) and it is useful to use the approach mentioned in Section~\ref{sec:sparse}. The ratings (linking attributes) are assumed Poisson distributed, due to the ordinal nature of the ratings going from 1-5. This is a sensible model choice in this context due to both the ordinal and discrete nature of the ratings. The clustering task will cluster groups of individuals giving a similar rating to specific groups of movies. One could expect that this would glean more information than treating the 1-5 ratings as purely categorical since the Poisson model captures in some ways the ordinal features of the data. A missing rating is just given the value 0 when assuming the ratings are Poisson. The data is shown in Figure~\ref{fig:ml100k}. One run of each of the algorithms was carried out each with two restarts with results in Table~\ref{tab:ml100k}. The reordered data is shown for algorithms A1 and A3 in Figure~\ref{fig:ml100k}.

\begin{table}
\begin{center}
\begin{tabular}{cccc}
\hline \hline
Algorithm & maximum ICL & run time (sec) & $(K,G)$\\
\hline
A0 & -646597.7  & 1098.529 & (55,64) \\
A1 & -646597.7  & 368.032 & (55,64)\\
A2 & -646268.2  & 525.818 & (56,62)\\
A3 & -646268.2  & 160.162 & (56,62)\\
\hline \hline
\end{tabular}
\end{center}
\caption{Results for one run of each of the four algorithms on Movie-Lens data with same random seed.} \label{tab:ml100k}
\end{table} 

A glance at these results reveals that the pruning based algorithms give different number of clusters to those that do not involve pruning. This could be due to the effect of the value of the difference in exact ICL considered for the pruning. A value of this threshold which is too high could introduce some degeneracy into the search space, with labels settling in sub-optimal groups. The result for this dataset can vary depending on the pruning threshold chosen. However, pruning does give a considerable reduction in computing time which is favourable from a scalability viewpoint. This could be seen as an option when the data size is such that a run of the full algorithm is just too expensive. 


%
%
%
%

\section{Discussion}

We have presented an approach to grouping or clustering subsets of nodes in bipartite networks which share similar linking properties. There are two main advantages to our work. The first is that  we model linking patterns between nodes in the two node sets in a statistical way. The second is that we provide a principled inference technique for the number of clusters in the node sets by appealing to a known information criterion due to~\citeasnoun{Biernacki01}. The fact that the exact ICL may be computed due to the tractability assumptions we make, means that it can be iteratively updated for each proposed new labelling of a node in either node set.  

The approach presented here can be considered favourable to the MCMC techniques of~\citeasnoun{Wyse12} for a few reasons. If only an optimal block structure is required our algorithm provides this in a more scalable framework than MCMC. Also, in summarizing MCMC output, often the maximum a posteriori (MAP) estimate is used as an overall summary, since it is not possible to average clusterings with a difference number of clusters. The concept of this MAP estimator is in somewhat equivalent to the configuration of labels and clusters that maximizes the ICL. Probably the most important reason is that MCMC is simply not a viable option for large bipartite networks due to the declining mixing performance observed and the long run times required in such cases. The simulation study shows that we gain as much information from the greedy algorithm as we would from a longer and more time consuming MCMC run, which is encouraging. This being said, there is always a risk that the ICL greedy algorithm we discuss could get trapped in a local maximum. Due to the fast run time, we can circumvent this somewhat in practice by running, say 10 instances of the greedy algorithm and taking the one that gives the highest ICL (indicating the best clustering). The greedy algorithm (especially the sparse versions) scale well enough to make this feasible.  The risk that this will happen to the MCMC algorithm is less due to its stochastic nature. ICL does still possess many of the benefits of MCMC, the strongest of which is that it allows for automatic choice of the number of clusters. Considering the complexity of the task it performs (model selection and clustering), the ICL algorithm does appear to scale quite well, running on the Movie-Lens 100k data in less than 3 minutes, and the congressional voting data in less than a second. For many of the applications of bi-clustering in the literature, this is competitive. In addition, we provide code (written in C with an R wrapper) which is available from the first author's webpage 
\begin{verbatim}
https://sites.google.com/site/jsnwyse/   
\end{verbatim}
(a Makefile is provided to compile on Unix-alikes). 

In Section~\ref{sec:sec4} we showed how sparsity (a common feature of observed networks) can be exploited to provide a faster version of the greedy search exact ICL algorithm. Although not possible to provide exact quantification for the order of speed up (in terms of $N$ and $M$) we observed considerable computational gain in exploiting sparsity. The applicability of these ideas may stretch well beyond implementation of the LBM. For example, it may be possible to use the same kind of representation for SBMs. 

We believe that the latent blockmodel is an appealing model for bipartite networks, although to our knowledge, no authors have appeared to explicitly discuss it in the context of network modelling. One drawback of the LBM in its original form was the lack of any principled information criterion (such as Bayesian Information Criterion) for choosing the number of clusters. However, the work of~\citeasnoun{Keribin13} and the ideas presented in this paper overcome this issue. For this reason, the LBM can now be seen as a rich model for which there is capacity to model group structure in bipartite networks. The approach we take also means it is scalable for larger networks where sparsity can be exploited. 

%

\bibliography{bibliography}

\end{document}